\pdfoutput=1
\documentclass[a4paper,11pt]{article}
\usepackage{pos,orcidlink,multicol}
\hypersetup{
 pdfauthor={Florian Ecker, Daniel Grumiller, and Patricio Salgado-Rebolledo}, linkcolor=blue,urlcolor=magenta,filecolor=green,citecolor=red,pdfstartview=FitV,bookmarksopen=true}

\newcommand{\mytitle}{Postcarrollian gravity}

\DeclareMathOperator{\extdm}{d}

\newcommand{\extd}{\extdm \!}
\newcommand*{\dd}{\extd}

\newcommand{\XM}{X_{\textrm{\tiny M}}}
\newcommand{\XH}{X_{\textrm{\tiny H}}}
\newcommand{\XP}{X_{\textrm{\tiny P}}}
\newcommand{\XZ}{X_{\textrm{\tiny Z}}}

\newcommand{\aM}{a_{\textrm{\tiny M}}}
\newcommand{\aH}{a_{\textrm{\tiny H}}}
\newcommand{\aP}{a_{\textrm{\tiny P}}}
\newcommand{\aZ}{a_{\textrm{\tiny Z}}}
\newcommand{\aB}{a_{\textrm{\tiny B}}}
\newcommand{\aG}{a_{\textrm{\tiny G}}}
\newcommand{\aS}{a_{\textrm{\tiny S}}}
\newcommand{\aC}{a_{\textrm{\tiny C}}}
\newcommand{\aL}{a_{\textrm{\tiny L}}}

\newcommand{\xM}{x_{\textrm{\tiny M}}}
\newcommand{\xH}{x_{\textrm{\tiny H}}}
\newcommand{\xP}{x_{\textrm{\tiny P}}}
\newcommand{\xZ}{x_{\textrm{\tiny Z}}}
\newcommand{\xB}{x_{\textrm{\tiny B}}}

\newcommand{\epsM}{\varepsilon_{\textrm{\tiny M}}}
\newcommand{\epsH}{\varepsilon_{\textrm{\tiny H}}}
\newcommand{\epsP}{\varepsilon_{\textrm{\tiny P}}}
\newcommand{\epsZ}{\varepsilon_{\textrm{\tiny Z}}}
\newcommand{\epsB}{\varepsilon_{\textrm{\tiny B}}}
\newcommand{\epsG}{\varepsilon_{\textrm{\tiny G}}}
\newcommand{\epsC}{\varepsilon_{\textrm{\tiny C}}}
\newcommand{\epsJ}{\varepsilon_{\textrm{\tiny J}}}
\newcommand{\epsS}{\varepsilon_{\textrm{\tiny S}}}

\newcommand{\0}{\underline{0}}
\newcommand{\1}{\underline{1}}
\newcommand{\2}{\underline{2}}

\newcommand{\eq}[2]{\begin{equation} #1 \label{#2} \end{equation}}

\numberwithin{equation}{section}

\title{\mytitle}

\author[a,b]{\orcidlink{0000-0002-0449-0081}Florian Ecker}
\author*[a,b]{\orcidlink{0000-0001-7980-5394}Daniel Grumiller}
\author[b,c,d]{\orcidlink{0000-0001-6832-6785}Patricio Salgado-Rebolledo}

\affiliation[a]{Institute for Theoretical Physics, TU Wien,\\
  Wiedner Hauptstrasse 8-10/136, A-1040 Vienna, Austria}

\affiliation[b]{Erwin Schr\"odinger International Institute for Mathematics and Physics,\\
Boltzmanngasse 9, A-1090 Vienna, Austria}

\affiliation[c]{Asia Pacific Center for Theoretical Physics (APCTP), Pohang, Gyeongbuk 37673, Korea}

\affiliation[d]{Instituto de Ciencias Exactas y Naturales (ICEN), Universidad Arturo Prat,\\
Playa Brava 3256, 1111346 Iquique, Chile}

\emailAdd{fecker@hep.itp.tuwien.ac.at}
\emailAdd{grumil@hep.itp.tuwien.ac.at}
\emailAdd{salgado.rebolledo@apctp.org}

\abstract{
We construct postcarrollian gravity models in two, three, and four spacetime dimensions by applying algebraic expansion methods. As a byproduct, we present the most general postcarrollian 2d dilaton gravity model, construct its solutions and discuss some boundary aspects, including Schwarzian-type boundary actions. In 3d, we propose Brown--Henneaux-like boundary conditions, generalizing a corresponding Carrollian analysis, and derive the postcarrollian asymptotic symmetry algebra with its central extensions.
}

\FullConference{Proceedings of the Corfu Summer Institute 2024 ``School and Workshops on Elementary Particle Physics and Gravity'' (CORFU2024)\
12 - 26 May, and 25 August - 27 September, 2024\\
Corfu, Greece\\}

\tableofcontents

\begin{document}
\maketitle


\section{Introduction}\label{se:1}

Carroll symmetries are spacetime symmetries dual to Galilean symmetries, leading to absolute space and relative time. Carroll boosts leave invariant space but linearly transform time, $t\to t+b_i x^i$, where $b_i$ are the Carroll boost parameters. The Carroll algebra infinitesimally generating these symmetries formally emerges as the limit of vanishing speed of light ($c\to 0$) from an \.In\"on\"u--Wigner contraction of the Poincar\'e algebra \cite{LevyLeblond1965,Gupta1966}.

A distinguishing feature of the Carroll algebra (as opposed to the Poincar\'e or Galilei algebras) is that the Hamiltonian $H$ commutes with the boost generators $B_a$,
\eq{
[H,\,B_a]=0\,.
}{eq:intro1}
Physically, this is remarkable since it implies that Carroll boosting any state does not change its energy, in contrast to our relativistic and non-relativistic intuition. The vanishing commutator \eqref{eq:intro1} is reflected in physical properties of systems with Carroll symmetries, such as pp-waves \cite{Souriau:1973,Elbistan:2023qbp}, zero-signature spacetimes \cite{Henneaux:1979vn}, cosmology \cite{Henneaux:1982qpq,Damour:2002et,deBoer:2021jej}, tachyon condensates \cite{Gibbons:2002tv}, flat space holography \cite{Barnich:2006av,Bagchi:2010zz,Bagchi:2012yk,Bagchi:2012xr,Barnich:2012xq,Barnich:2013yka,Duval:2014uva,Bagchi:2014iea,Donnay:2022aba,Bagchi:2022emh}, fluid-gravity correspondence \cite{deBoer:2017ing,Ciambelli:2018xat}, null hypersurfaces \cite{Penna:2018gfx,Donnay:2019jiz,Ciambelli:2019lap,Redondo-Yuste:2022czg,Freidel:2022vjq,Gray:2022svz}, null infinity \cite{Ciambelli:2018wre,Figueroa-OFarrill:2021sxz,Herfray:2021qmp,Mittal:2022ywl}, fractons \cite{Bidussi:2021nmp,Marsot:2022imf,Figueroa-OFarrill:2023vbj}, current-current deformations \cite{Rodriguez:2021tcz}, flat bands \cite{Bagchi:2022eui}, tensionless strings \cite{Bagchi:2015nca}, near horizon soft hair \cite{Bagchi:2022iqb}, near horizon strings \cite{Bagchi:2023cfp}, Carroll black holes \cite{Ecker:2023uwm}, spacetime subsystem symmetries \cite{Baig:2023yaz,Kasikci:2023tvs}, swiftons \cite{Ecker:2024czx}, open null strings \cite{Bagchi:2024qsb}, shallow water waves \cite{Bagchi:2024ikw}, phase separation \cite{Biswas:2025dte}, etc. (see \cite{deBoer:2023fnj,Ciambelli:2023xqk,Bergshoeff:2024ilz} for additional Refs.)

Gauging the Carroll algebra leads to Carroll gravity \cite{Hartong:2015xda}. Simple examples of Carroll gravity theories are Carroll dilaton gravity models in two dimensions (2d) \cite{Grumiller:2020elf,Gomis:2020wxp,Aviles:2022xyx} such as the Carroll Jackiw--Teitelboim (JT) model or the Carroll Callan--Giddings--Harvey--Strominger (CGHS) model. The key property \eqref{eq:intro1} holds for all these models.

In some applications, it can be of interest to deviate from the Carroll limit without fully restoring the Poincar\'e symmetries. The purpose of this work is to do this in a systematic way in the context of Carroll gravity. Thus, our main goal is to construct postcarrollian gravity theories. To achieve this goal, we focus on the algebraic perspective, i.e., we allow the commutator \eqref{eq:intro1} to be non-zero but in a minimalistic way to make sure we are not fully back to Poincar\'e symmetries. In the remainder of this introduction, we make this notion precise. 

An expansion for small speed of light was previously considered in \cite{Hansen:2021fxi} using the second-order formulation of general relativity. In that case, the leading order theory is of ``electric'' character and the corrections refer to that theory. In this work, by contrast, we work in the first-order formulation, which is known to produce the ``magnetic'' theory at leading order. Thus, the postcarrollian expansion we perform in this work has to be thought of as introducing corrections to the magnetic Carroll theory. 

Before addressing the general situation in arbitrary dimensions, we glean some intuition from 2d, where the commutator \eqref{eq:intro1} simplifies to $[H,\,B]=0$. From a purely algebraic perspective, we can make this commutator less Carrollian by relaxing the condition that its right-hand side vanishes. Instead, we allow the appearance of a new generator, $M$, but then demand that this new generator be central. The commutation relations $[H,\,B]=M$ and $[M,\,\cdot]=0$ are one step closer to the Poincar\'e algebra, in the sense that the Hamiltonian and boost generators no longer commute with each other, but instead, they commute into a generator that commutes with everything. 

While some aspects of this intuition remain apposite in higher dimensions, the analog $M_a$ can no longer be central, simply because it carries a spatial vector index and thus must not commute trivially with rotations. We explain now how this works in detail.

The (A)dS Carroll algebra can be extended by allowing postcarrollian corrections in the relation between the (A)dS gravitational fields and the Carrollian fields. We denote the Lorentzian vielbein by $E^A_\mu =(E^0_\mu, E^a_\mu)$ and the spin connection by $\Omega^{AB}_\mu =(\Omega^{0a}_\mu, \Omega^{ab}_\mu)$. The Carrollian fields consist of the clock 1-form $\tau_\mu$, the spatial vielbein $e^a_\mu$, the Carroll boost connection $\omega^a_\mu$, and the spatial rotation connection $\omega^{ab}_\mu$. Here, $A$, $B$ denote anholonomic spacetime indices, $a$, $b$ anholonomic spatial indices and $0$ an anholonomic time index. The holonomic components are denoted by $\mu $ for spacetime indices and $i$, $j$ for spatial indices. The expansion we perform reads
\begin{equation}\label{expansionRelFields}\begin{aligned}
E^0_\mu&= c \,\tau_\mu + c^3\, l_\mu & \qquad
E^a_\mu&=e^a_\mu+ c^2\, m^a_\mu \\
\Omega^{0a}_\mu&= c \,\omega^a_\mu + c^3\, \sigma^a_\mu & \qquad
\Omega^{ab}&= \omega^{ab}_\mu + c^2\, \sigma^{ab}_\mu
\end{aligned}\end{equation}
where $c$ is our expansion parameter and all the relations are supplemented by corrections. Terms of $\mathcal{O}(c^4)$ have been neglected. By introducing generators of Lorentz transformations $\mathcal J_{AB}$ and (A)dS translations $\mathcal P_A$, an (A)dS valued gauge connection 1-form
\begin{equation}\label{connectionAds}\begin{aligned}
A&= E^A \, \otimes \mathcal P_A+ \frac12 \Omega^{AB} \,\otimes  \mathcal J_{AB}
\end{aligned}\end{equation}
can be expanded around the Carroll limit, $c\to 0$, by inserting \eqref{expansionRelFields} into \eqref{connectionAds} and subsequently renaming
\begin{equation}\label{expgenAdSCarroll}\begin{aligned}
H &= c \otimes \mathcal P_0 &\qquad P_a &= c^0 \otimes \mathcal P_a & \qquad B_a &= c \otimes \mathcal J_{0a} &\qquad J_{ab} &=  c^0 \otimes \mathcal J_{ab}
\\
L &= c^3 \otimes\mathcal P_0 &\qquad M_a &= c^2 \otimes \mathcal P_a & \qquad
K_a &= c^3 \otimes \mathcal J_{0a} &\qquad S_{ab} &= c^2 \otimes \mathcal J_{ab} ~.
\end{aligned}\end{equation}
The construction \eqref{expgenAdSCarroll} is a particular case of a Lie algebra expansion \cite{Boulanger:2002bt,deAzcarraga:2002xi,Izaurieta:2006zz,Bergshoeff:2019ctr} and can be obtained as a particular quotient of an infinite-dimensional extension of the (A)dS Carroll algebra \cite{Gomis:2019nih}. 
The first row of generators closes into the standard (A)dS Carroll algebra \cite{Bacry:1968zf,Matulich:2019cdo} in the limit $c\rightarrow 0$, and is associated with the leading order in the gauge field expansion \eqref{expansionRelFields}. The second row of generators is identified as the postcarrollian corrections of the (A)dS Carroll generators.

After these definitions, the postcarrollian gauge field components read
\begin{equation}\label{connectionA}
A_\mu=\tau_\mu\, H + e^a_\mu\, P_a + \omega^a_\mu\, B_a + \frac12 \omega^{ab}_\mu\, J_{ab}
+l_\mu\, L + m^a_\mu\, M_a + \sigma^a_\mu\, K_a + \frac12 \sigma^{ab}_\mu\, S_{ab}\
\end{equation}
 and the postcarrollian generators \eqref{expgenAdSCarroll} satisfy the commutation relations\footnote{%
 If we set the cosmological constant to zero, $\Lambda=0$, the algebra \eqref{expgenAdSCarroll} describes postcarrollian corrections to the flat Carroll symmetries. In this case, additionally the commutators \eqref{eq:1}, \eqref{eq:0}, \eqref{eq:2}, and \eqref{eq:3} vanish.}
\begin{multicols}{2}
\begin{subequations}\label{PostCalgebra}
\setlength{\abovedisplayskip}{-14pt}
\allowdisplaybreaks
\begin{align}
    [J_{ab},J_{cd}]&=4\delta _{[a[c}J_{d]b]}\\
    [J_{ab},P_c]&=2\delta_{c[b}P_{a]}\\
    [J_{ab},B_c]&=2\delta_{c[b}B_{a]}\\
    [P_a,P_b]&=-\Lambda\,J_{ab} \label{eq:1}\\
    [B_a,P_b]&=\delta_{ab}\,H\\
    [P_a,H]&=\Lambda\,B_a \label{eq:0}\\
    [B_a,B_b]&=S_{ab}\\
    [B_a,H]&=M_a\\
    [J_{ab},S_{cd}]&=4\delta_{[a[c}S_{d]b]}\\
    [J_{ab},M_c]&=2\delta_{c[b}M_{a]}=[S_{ab},P_c]\\
    [J_{ab},K_c]&=2\delta_{c[b}K_{a]}=[S_{ab},B_c]\\
    [P_a,M_b]&=-\Lambda\,S_{ab}\label{eq:2}\\
    [K_a,P_b]&=\delta_{ab}\,L=[B_a,M_b]\\
    [M_a,H]&=\Lambda\,K_a=[P_a,L]\label{eq:3}\\
    \nonumber\\
    & \!\!\!\!\!\!\!\!\!\!\!\!\!\!\!\! \textrm{commutators\;not\;displayed\;vanish.} \nonumber
\end{align}
\end{subequations}
\end{multicols}
\noindent The commutators in the left half form the Carroll algebra provided we set to zero the right-hand sides in the last two lines. The commutators in the right half all involve one subleading generator on each side of the equalities that mirror the corresponding leading generators from the left half.

Gravitational theories that admit a gauge formulation in terms of Lie algebra-valued fields, such as \eqref{connectionA}, allow for a postcarrollian expansion that is equivalently obtained by expanding either the (A)dS Carroll generators or the corresponding gauge fields. On the other hand, the expansion \eqref{expansionRelFields} allows to construct postcarrollian gravitational models even in the case that a formulation in terms of Lie algebra-valued objects is not available. Examples of the former case include Chern--Simons and BF theories, whereas the latter situation arises in deformed BF models and Macdowell--Mansouri-type Lagrangians.

This paper is organized as follows. In section \ref{se:2}, we start with postcarrollian BF theories, with a focus on the postcarrollian JT and CGHS models. In section \ref{se:3}, we generalize to generic postcarrollian 2d dilaton gravity models. In section \ref{se:4}, we construct all solutions for all these models. In section \ref{se:5}, we focus on boundary aspects, propose specific boundary conditions, and discuss Schwarzian-type boundary actions for specific postcarrollian 2d dilaton gravity models. In section \ref{se:6}, we generalize to postcarrollian gravity in three dimensions (3d) and four dimensions (4d), including an asymptotic symmetry discussion for Brown--Henneaux-like boundary conditions in 3d. In section \ref{se:7}, we conclude with an outlook.


\section{Postcarrollian BF gravity theories in 2d}\label{se:2}

In this section, we construct simple postcarrollian gravitational theories in 2d. Consider a family of BF-models 
\eq{
    I_{\textrm{\tiny BF}}[\mathcal{X},A]=\frac{k}{2\pi}\int \langle \mathcal{X},F\rangle 
}{eq:BF_model}
based on a Lie algebra $\mathfrak{g}$ with an invariant metric $\langle \cdot ,\cdot \rangle :\mathfrak{g}\times \mathfrak{g}\to \mathbb{R}$. While generalizations to Lie algebras without such a metric are possible (see, e.g., \cite{Grumiller:2020elf} and section \ref{se:3}), we restrict to this case in the current section. The field $A=A_\mu ^I\mathbf{e}_I\dd x^\mu $ is a $\mathfrak{g}$-valued one-form and $\mathcal{X}=\mathcal{X}^I\mathbf{e}_I$ is a $\mathfrak{g}$-valued scalar where we chose a particular basis $\{\mathbf{e}_I\}$ for $\mathfrak{g}$ satisfying $[\mathbf{e}_I,\mathbf{e}_J]=c_{IJ}{}^K\mathbf{e}_K$ with structure constants $c_{IJ}{}^K$. From $A$ we define the curvature two-form $F= \dd A + A \wedge A$ and for gauge parameters $\lambda =\lambda ^I\mathbf{e}_I$, one can show that the action is invariant under the gauge variations
\begin{align}\label{eq:gauge_trafos}
    \delta _\lambda \mathcal{X}&=[\mathcal{X},\lambda ] & \delta _\lambda A&=\dd \lambda +[A,\lambda ] ~.
\end{align}
The finite versions of the gauge transformations are obtained by integrating $\lambda$ to the group element $g=\exp(\lambda)$ and act as
\begin{align}
    \mathcal{X}&\to g^{-1}\mathcal{X}g & A&\to g^{-1}\big(\dd +A\big)g ~.
\end{align}

The equations of motion 
\begin{align}\label{eq:BF_EOM}
    F&=0 & \dd \mathcal{X}+[A,\mathcal{X}]&=0 
\end{align}
make it evident that on-shell the field $A$ is locally pure gauge, $A=g^{-1}\dd g$, while $\mathcal{X}$ is an element of the stabilizer for this configuration, $\delta _{\mathcal{X}}A=0$. We shall also need the quadratic Casimir, 
\begin{align}\label{eq:BF_Casimir}
    \mathcal{C}=\langle \chi ,\chi \rangle
\end{align}
another gauge-invariant quantity that is moreover conserved on-shell, $\dd \mathcal{C}=0$.

So far, these models are just gauge theories on a 2d manifold. To identify them as gravitational theories, we have to relate the field variables to geometric variables once we choose an algebra $\mathfrak{g}$ together with an invariant metric. We shall consider five-dimensional algebras $\mathfrak{g}=\{B,H,P,M,Z\}$ here and choose the gauge field $A$ as 
\eq{
    A=\omega B+\tau H+eP+mM+zZ
}{eq:ladida}
where $e$ is the spatial einbein, $\tau$ is the clock one-form, $\omega$ is the boost connection, and $m$ is a one-form we shall interpret as a postcarrollian correction to $e$. The generator $Z$ is a central term that we need to obtain a non-degenerate bilinear form in the case of the postcarrollian CGHS model; correspondingly, $z$ is an abelian 1-form that goes along for the ride. This is in analogy to a construction already used in \cite{Cangemi:1992bj,Afshar:2019axx} for the case of the Poincaré algebra in 2d. In terms of the gauge field variables, the Carrollian metric is given by $h_{\mu \nu}=e_\mu e_\nu$ and the Carrollian vector field $v^\mu $ is determined by $v^\mu \tau_\mu =-1$, $v^\mu e_\mu =0$. 

For writing down the action it is convenient to express the field $\mathcal{X}$ in terms of its dual version $\mathcal{X}^\ast =\langle \mathcal{X},\cdot \rangle =\mathcal{X}^\ast _IE^I$ where $E^I(e_J)=\delta ^I_J$. Explicitly, we write
\eq{
    \mathcal{X}^\ast =X B^\ast +\XH H^\ast + \XP P^\ast +\XM M^\ast + \XZ Z^\ast 
}{eq:x}
with the dual algebra basis elements satisfying $B^\ast (B)=H^\ast (H)=P^\ast (P)=M^\ast (M)=Z^\ast (Z)=1$. Under a gauge transformation $\lambda $ this field transforms coadjointly as $\delta _\lambda \mathcal{X}^\ast (\cdot )=-\mathcal{X}^\ast ([\cdot ,\lambda ])$. We relate the gauge transformations to spacetime diffeomorphisms generated by a vector field $\xi $ and internal transformations generated by $\{B,M,Z\}$ by defining, without loss of generality
\begin{align}
    \lambda &=\xi ^\mu A_\mu +\Bar{\lambda} & \Bar{\lambda}&=\lambda _{\textrm{\tiny B}}B+\lambda _{\textrm{\tiny M}}M+\lambda _{\textrm{\tiny Z}}Z
\end{align}
such that ($\mathcal{L}_\xi$ denotes Lie derivatives along $\xi$ and $i_\xi$ contraction with $\xi$ from the left)
\begin{align}
    \delta _\lambda A&=\delta _{\Bar{\lambda}}A+\mathcal{L}_\xi A+i_\xi F &
    \delta _\lambda \mathcal{X}&=\delta _{\Bar{\lambda}}\mathcal{X}+\mathcal{L}_\xi \mathcal{X}-i_\xi \big(\dd \mathcal{X}+[A,\mathcal{X}]\big) ~.
\end{align}
Since the last terms vanish once we evaluate on the equations of motion \eqref{eq:BF_EOM}, we see that $\xi$ indeed generates diffeomorphisms on-shell.

\subsection{Postcarrollian JT model}\label{se:2.1}

Our first example starts with the postcarrollian algebra
\eq{
    [B,P]=H\qquad\qquad[B,H]=M\qquad\qquad[H,\,P]=-\Lambda B 
}{eq:angelinajolie}
with cosmological constant $\Lambda $. The $\mathfrak{u}(1)$-sector associated to $Z$ does not feature in these brackets and could therefore be truncated away. We will, however, keep it for now since it will play a role in taking the limit $\Lambda\to 0$. An invariant metric on $\mathfrak{g}$ is given by
\eq{
    \langle H,\,H\rangle=\mu \Lambda \qquad\qquad \langle M,\,P\rangle=-\mu \Lambda \qquad\qquad \langle B,\,B\rangle =\mu \qquad\qquad \langle Z,Z \rangle =\nu 
}{eq:wtf}
where we are free to choose the two real constants $\mu,\nu\neq 0$. 

Without the $Z$-sector, the algebra \eqref{eq:angelinajolie} was already studied in \cite{Grumiller:2020elf,Gomis:2020wxp} where the central extension $M$ was introduced in order to obtain a non-degenerate bilinear form. Writing out the BF-action \eqref{eq:BF_model} for these choices, we obtain the postcarrollian JT-model
\eq{
    I_{\textrm{\tiny JT}}=\frac{k}{2\pi}\int \big[X\,\extd\omega+\XH\,\big(\extd\tau+\omega\wedge e\big)+\XP\,\extd e+\XM\,\big(\extd m+\omega\wedge\tau\big)-\Lambda X\tau \wedge e+\XZ\,\extd z \big]~.
}{eq:JT_action}
The equations of motion \eqref{eq:BF_EOM} for this model read explicitly
\begin{subequations}
    \label{eq:EOM_JT}
\begin{align}
  \delta X &&  \textrm{Constant\;Carroll\;curvature:} && &\Omega =\extd\omega = \Lambda \tau \wedge e \\
  \delta \XH && \textrm{No\;Carroll\;torsion:} && &T =\extd\tau + \omega\wedge e = 0 \\
  \delta \XP && \textrm{No\;intrinsic\;torsion:} && &\Theta = \extd e = 0 \\
  \delta \XM && \textrm{No\;postcarrollian\;torsion:} && & \Xi = \extd m + \omega\wedge\tau = 0 \\
  \delta \XZ && \textrm{Vanishing\;electric\;field:} && &E = \extd z = 0 \\
  \delta \omega && \textrm{Carroll\;metric:} && &\extd X + \XH\,e+\XM\,\tau = 0 \\
  \delta\tau && \textrm{Carroll\;Casimir:} && & \extd\XH-\XM\,\omega-\Lambda Xe = 0 \\
  \delta e && \textrm{Auxiliary\;field:} && &\extd\XP - \XH\,\omega +\Lambda X\,\tau = 0 \\
  \delta m && \textrm{Abelian\;Casimir\;I:} && & \extd\XM=0\\
  \delta z && \textrm{Abelian\;Casimir\;II:} && & \extd\XZ=0 ~.
\end{align}
\end{subequations}

We derive the general solution to these equations in section \ref{se:4}. The addition of the generator $M$ makes it possible to solve for the boost connection $\omega$ uniquely in terms of the fields $e$, $\tau$, and $m$. This is in contrast to the Carrollian case investigated in \cite{Ecker:2023uwm} and more generally in \cite{Bergshoeff:2017btm,Campoleoni:2022ebj} where a part of the boost connection stays undetermined on-shell. We also highlight the constant curvature $\Omega =\Lambda \tau \wedge e$ of all solutions, which is a characteristic feature of JT gravity. The action of internal gauge transformations generated by $\Bar{\lambda} =\lambda _{\textrm{\tiny B}}B+\lambda _{\textrm{\tiny M}}M+\lambda _{\textrm{\tiny Z}}Z$ on the fields is given by
\begin{subequations}\label{eq:internal_gaugetrafos}
\begin{align}
    \delta _{\Bar{\lambda}} \tau &=-e\lambda _{\textrm{\tiny B}} & \delta _{\Bar{\lambda}}e&=0 & \delta _{\Bar{\lambda}}\omega &=\dd \lambda _{\textrm{\tiny B}} & \delta _{\Bar{\lambda}}m&=-\tau \lambda _{\textrm{\tiny B}} & \delta _{\Bar{\lambda}}z&=0\\
    \delta _{\Bar{\lambda}}\XH &=\lambda _{\textrm{\tiny B}}\XM & \delta _{\Bar{\lambda}}\XP &=\lambda _{\textrm{\tiny B}}\XH & \delta _{\Bar{\lambda}}X&=0 & \delta _{\Bar{\lambda}}\XM &=0 & \delta _{\Bar{\lambda}}\XZ &=0
\end{align}
\end{subequations}
where one can recognize local Carroll boosts generated by $\lambda_{\textrm{\tiny B}}$ (see, e.g., \cite{Ecker:2023uwm}). One can also check that besides the Carroll boost-invariant scalar fields $X,\XM, \XZ$ there is another boost-invariant combination, $\frac{1}{2}\XH ^2-\XP \XM$.

\subsection{Postcarrollian CGHS model}\label{se:2.2}

Starting from the postcarrollian JT model introduced in the previous subsection, we perform the limit $\Lambda \to 0$ to obtain a postcarrollian version of the CGHS model \cite{Callan:1992rs} in the formulation of Cangemi and Jackiw \cite{Cangemi:1992bj}. For this, we first redefine the generator for Carroll boosts by
\eq{
    B\to B+\frac{1}{\Lambda}Z
}{eq:lalapetz}
which is just a shift by a trivial central term. However, once the limit has been taken, this central term becomes non-trivial and therefore cannot be transformed away by a change of basis. The algebra reads
\eq{
    [B,\,P]=H\qquad\qquad[B,\,H]=M\qquad\qquad[H,\,P]=Z 
}{eq:CGHS_algebra}
and corresponds to the doubly extended Carroll algebra in Table 1 of \cite{Grumiller:2020elf}. The limit on the level of the bilinear form is performed by choosing $\mu =1/\Lambda$, $\nu =-\Lambda$, leading to
\begin{align}\label{eq:CGHS_bil}
    \langle H,\,H\rangle&=1  & \langle M,\,P\rangle &=-1 & \langle B,\,Z\rangle &=-1 
\end{align}
with all the other products being zero. Here, it becomes evident why the introduction of the additional $\mathfrak{u}(1)$-sector associated to $Z$ was necessary, as otherwise there was no way of making the bilinear form non-degenerate. This is in the spirit of the Poincaré case where a similar construction leads to the 2d Maxwell algebra in the limit $\Lambda \to 0$ \cite{Cangemi:1992bj,Afshar:2019axx}. The action of the postcarrollian CGHS model
\eq{
    I_{\textrm{\tiny CGHS}}=\frac{k}{2\pi}\int \big[X\,\extd\omega+\XH\,\big(\extd\tau+\omega\wedge e\big)+\XP\,\extd e+\XM\,\big(\extd m+\omega\wedge\tau\big)+\XZ\,\big(\extd z+\tau\wedge e\big)\big]
}{eq:CGHS_action}
produces the equations of motion
\begin{subequations}
    \label{eq:CGHS_EOM}
\begin{align}
  \delta X &&  \textrm{No\;Carroll\;curvature:} && &\Omega =\extd\omega = 0 \\
  \delta \XH && \textrm{No\;Carroll\;torsion:} && &T =\extd\tau + \omega\wedge e = 0 \\
  \delta \XP && \textrm{No\;intrinsic\;torsion:} && &\Theta = \extd e = 0 \\
  \delta \XM && \textrm{No\;postcarrollian\;torsion:} && & \Xi = \extd m + \omega\wedge\tau = 0 \\
  \delta \XZ && \textrm{Constant\;electric\;field:} && &E = \extd z = -\tau\wedge e \\
  \delta \omega && \textrm{Carroll\;metric:} && &\extd X + \XH\,e+\XM\,\tau = 0 \\
  \delta\tau && \textrm{Carroll\;Casimir:} && & \extd\XH-\XM\,\omega+\XZ\,e = 0 \\
  \delta e && \textrm{Auxiliary\;field:} && &\extd\XP - \XH\,\omega-\XZ\,\tau = 0 \\
  \delta m && \textrm{Abelian\;Casimir\;I:} && & \extd\XM=0\\
  \delta z && \textrm{Abelian\;Casimir\;II:} && & \extd\XZ=0 ~.
\end{align}
\end{subequations}
Similar to the Lorentzian counterpart, this model only admits locally flat solutions. The action of internal gauge transformations is the same as in \eqref{eq:internal_gaugetrafos}.

We shall come back to this model in section \ref{se:5} when addressing boundary conditions and boundary actions.


\section{Postcarrollian 2d dilaton gravity}\label{se:3}

In this section, we generalize the models of the previous section to the most general deformation thereof by analogy to the Lorentzian case \cite{Grumiller:2021cwg}. This deforms the gauge symmetries of a theory but preserves the number of local physical and gauge degrees of freedom. Additionally preserving the dictionary between the gauge theoretic variables and the geometric variables, we obtain general theories of postcarrollian dilaton gravity. These models can be embedded in a much larger class of non-linear gauge theories \cite{Ikeda:1993fh} known as Poisson--Sigma models (PSM) \cite{Schaller:1994es}. They are given by actions of the form
\eq{
    I_{\textrm{\tiny PSM}}[X^I,\mathcal{A}_I]=\frac{k}{2\pi}\int \big[X^I\dd \mathcal{A}_I+\frac{1}{2}P^{IJ}\mathcal{A}_I\wedge \mathcal{A}_J\big]
}{eq:PSM_action}
with the target space being a Poisson manifold with Poisson tensor $P^{IJ}(X^K)$ and target space coordinates $X^I$. A consistent Poisson structure requires the non-linear Jacobi identities
\eq{
    P^{IL}\partial_LP^{JK} + P^{JL}\partial_LP^{KI} + P^{KL}\partial_LP^{IJ}= 0 ~.
}{eq:jacobi}
The fields $\mathcal{A}_I$ are one-forms on the base manifold and one-forms on the target space. 

With these properties, one can show that, up to boundary terms, the action \eqref{eq:action} is invariant under gauge transformations generated by parameters $\lambda _I=(\lambda _{\textrm{\tiny B}},\lambda _{\textrm{\tiny H}},\lambda _{\textrm{\tiny P}},\lambda _{\textrm{\tiny M}},\lambda _{\textrm{\tiny Z}})_I$
\begin{align}
    \delta _\lambda X^I&=\lambda_J\,P^{JI} & \delta _\lambda \mathcal{A}_I&=\dd \lambda _I+\partial _IP^{JK}\mathcal{A}_J\lambda _K ~.
\end{align}
If the Poisson tensor is linear in the target space coordinates, then the gauge field transforms like in Yang--Mills or BF theory (see section \ref{se:2}); otherwise, we have non-linear gauge symmetries. 

\subsection{Postcarrollian dilaton gravity action and equations of motion}

To make contact with the postcarrollian gravity models from before, we identify the fields as 
\begin{align}
    X^I & =(X, \XH, \XP, \XM, \XZ )^I & \mathcal{A}_I&=(\omega ,\tau ,e,m,z)_I
\end{align}
and choose the Poisson tensor 
\begin{align}\label{eq:P}
    P^{IJ} =\begin{pmatrix}
    0 & \XM & \XH & 0 & 0 \\
    -\XM  &   0 & \mathcal{V} & 0 & 0 \\
     -\XH &  -\mathcal{V}   &   0 & 0 & 0 \\
     0 &  0   &  0   & 0 & 0 \\
     0 &  0   &   0  &  0 & 0
\end{pmatrix} ^{IJ}
\end{align}
where
\eq{
\mathcal{V}=\mathcal{V}(X,\XM,\XZ,\XH ^2-2\XP\XM)
}{eq:nolabel}
is any function of Carroll boost invariant arguments. For the special choices $\mathcal{V}=\XZ $ we recover the postcarrollian CGHS model \eqref{eq:CGHS_action} while for $\mathcal{V}=\Lambda X$ we get the postcarrollian JT model \eqref{eq:JT_action}. 

Using BRST arguments, it can be shown along the lines of \cite{Grumiller:2021cwg} that models with Poisson tensor \eqref{eq:P} are the most general deformation. The equations of motion for these models are richer as compared to Carrollian 2d dilaton gravity, cf.~Eqs.~(8) in \cite{Ecker:2023uwm}.
\begin{subequations}
    \label{eq:EOM_PSM}
\begin{align}
  \delta X &&  \textrm{Carroll\;curvature:} && &\Omega =\extd\omega = -\partial_X{\cal V}\,\tau\wedge e \\
  \delta \XH && \textrm{Carroll\;torsion:} && &T =\extd\tau + \omega\wedge e = -\partial_{\XH}{\cal V}\,\tau\wedge e  \\
  \delta \XP && \textrm{Intrinsic\;torsion:} && &\Theta = \extd e = -\partial_{\XP}{\cal V}\,\tau\wedge e \label{eq:torsion_eq}\\
  \delta \XM && \textrm{Postcarrollian\;torsion:} && & \Xi = \extd m + \omega\wedge\tau = -\partial_{\XM}{\cal V}\,\tau\wedge e \label{eq:postcalltorsioneq}\\
  \delta \XZ && \textrm{Electric\;field:} && &E = \extd z = -\partial_{\XZ}{\cal V}\,\tau\wedge e \\
  \delta \omega && \textrm{Carroll\;metric:} && &\extd X + \XH\,e+\XM\,\tau = 0 \label{eq:carr_metric_eq}\\
  \delta\tau && \textrm{Carroll\;Casimir:} && & \extd\XH-\XM\,\omega+{\cal V}\,e = 0 \\
  \delta e && \textrm{Auxiliary\;field:} && &\extd\XP - \XH\,\omega-{\cal V}\,\tau = 0 \\
  \delta m && \textrm{Abelian\;Casimir\;I:} && & \extd\XM=0\label{eq:CasXM}\\
  \delta a && \textrm{Abelian\;Casimir\;II:} && & \extd\XZ=0
\end{align}
\end{subequations}
In particular, one can see that the Carroll torsion equation in general implies non-vanishing torsion of $\omega$. For our purposes it will be enough to restrict $\mathcal{V}$ to the form\footnote{%
We could allow additionally arbitrary $\XM$-dependence in the potential. However, since on-shell $\XM$ is constant, this generalization almost does not change anything. Its only effect is to alter the postcarrollian torsion equation, which obtains new sources.
}
\eq{
    \mathcal{V}=V(X) - \big(\tfrac12\,\XH^2-\XM\XP\big)\,U(X)
}{eq:UV_family}
with two potential functions $V(X)$ and $U(X)$. This is inspired by the $UV$-family of 2d dilaton gravity \cite{Grumiller:2002nm}. 

Since the Poisson tensor is just another way of writing a Poisson bracket structure on the target manifold, we have for any two functions $\{F,\,G\}=P^{IJ}\partial_IF\partial_JG$, implying the Jacobi identities \eqref{eq:jacobi}. For the case \eqref{eq:UV_family} we shall see in section \ref{se:4} that there exist three Casimir functions $\mathcal{C}_i$ satisfying $\{\mathcal{C}_i,\cdot \}=0$ which corresponds to the Poisson tensor being rank $2$. Two of them are just $\XM$ and $\XZ$, see \eqref{eq:P}.

In summary, the generic postcarrollian 2d dilaton gravity bulk action is given by
\eq{
I_{\textrm{\tiny PC}} = \int\big[X\dd \omega +\XH \big(\dd \tau +\omega \wedge e\big)+\XP \dd e+\XM \big(\dd m +\omega \wedge \tau \big)+\XZ\,\extd z + \mathcal{V}\,\tau\wedge e\big]
}{eq:action}
with a generic potential of the form \eqref{eq:nolabel} that we shall restrict to \eqref{eq:UV_family} in this work. Compared to Carroll dilaton gravity \cite{Grumiller:2020elf,Ecker:2023uwm}, the main change in the action \eqref{eq:action} --- apart from the presence of additional subleading fields --- is the presence of a term proportional to $\omega\wedge\tau$. 

\subsection{Postcarrollian dilaton gravity as a limit}\label{se:3.1}

Before delving into a derivation of the general solutions to the models \eqref{eq:action} with \eqref{eq:UV_family} we show how they are obtained as a limit from Lorentzian theories. This motivates the interpretation as postcarrollian gravity theories, along the lines of our introduction. Following \cite{Grumiller:2002nm}, the Lorentzian first-order dilaton gravity action for $UV$-models is given by
\eq{
    I_{\textrm{\tiny Lorentz}}=\frac{\kappa}{2\pi}\int \big[X\dd \varpi +X_a\big(\dd E^a+\varpi \wedge E^b \epsilon ^a{}_b\big)-\Big(V(X)+\frac{U(X)}{2}X^cX_c\Big)\frac{1}{2}\epsilon _{ab}E^a\wedge E^b \big]
}{eq:yetanother}
where $\epsilon ^{01}=1$ and $X^cX_c=X^cX^d\eta _{cd}=-(X_0)^2+(X_1)^2$ with $\eta _{00}=-1=-\eta _{11}$. The fields $E^a$ denote the Lorentzian zweibein and $\varpi $ is the dual of the spin connection $\varpi ^a{}_b=\epsilon ^a{}_b\varpi $. These models can also be written as PSMs with a three-dimensional target space spanned by $(X,X^a)$.

Let us first look at how to perform the limit without postcarrollian corrections. We change the variables to
\begin{align}
    E^0&=c\tau & E^1&=e & \varpi &=c\omega \\
    X&=X & X_0&=\XH & X_1&=c\XP
\end{align}
such that we get
\eq{
    I_{\textrm{\tiny Lorentz}}=\frac{c\kappa}{2\pi}\int \Big[X\dd \omega +\XH (\dd \tau +\omega \wedge e)+\XP \dd e+\Big(V-\frac{U}{2}\XH ^2\Big)\tau \wedge e\Big]
    +\mathcal{O}(c^3)
}{eq:Lorentz}
which after rescaling the coupling constant $\kappa \to c^{-1}k$ and taking the limit $c\to 0$ is the magnetic Carroll dilaton gravity action \cite{Grumiller:2020elf,Ecker:2023uwm}. 

In order to get the postcarrollian action, one needs to introduce two additional Abelian BF-sectors before doing the expansion (this is analogous to similar constructions in 3d, see \cite{Bergshoeff:2016soe}). The extended Lorentzian action then reads
\begin{align}
    \Bar{I}_{\textrm{\tiny Lorentz}}&= I_{\textrm{\tiny Lorentz}}+\int \Big[\Phi \dd \eta +\Psi \dd \zeta +\frac{U(X)}{4}\Phi ^2 \epsilon _{ab}E^a\wedge E^b \Big] 
\end{align}
where $\eta $, $\zeta $ are a one-forms and $\Phi$, $\Psi $ are scalar fields. This can also be written as an extended PSM with a five-dimensional target space
\eq{
    \Bar{I}_{\textrm{\tiny Lorentz}}=\frac{\kappa}{2\pi}\int \big[X ^I\dd \mathcal{A}_I+\frac{1}{2}P^{IJ}\mathcal{A}_I\wedge \mathcal{A}_J\big]
}{eq:mystrangeaddiction}
with $\mathcal{A}_I=(\varpi ,E^0,E^1,\eta,\zeta)$, $X^I=(X,X^0,X^1,\Phi,\Psi)$, $\Bar{\mathcal{V}}=V(X)+\frac{U}{2}\big(X^aX_a-\Phi ^2\big)$, and
\begin{align}
P^{IJ} =\begin{pmatrix}
    0 & X^1 & -X^0 & 0 & 0  \\
    -X^1  &   0 &  \Bar{\mathcal{V}} & 0 & 0 \\
     X^0 &  -\Bar{\mathcal{V}}   &   0 & 0  & 0\\
     0 &  0   &  0   & 0  & 0 \\
    0  &  0   &  0   &  0  & 0
\end{pmatrix}^{IJ} 
\end{align}
satisfying the non-linear Jacobi identities \eqref{eq:jacobi}. 

We again make a change of variables that prepares us for the limit,
\begin{align}
    E^0&=c\tau & E^1&=e+\frac{c^2}{2}m & \varpi &=c\omega \\
    \eta &=e-\frac{c^2}{2}m & \zeta &=cz & \Psi &=\XZ \\
    X_0&=\XH & X_1&=\frac{1}{c}\XM +\frac{c}{2}\XP & \Phi &=-\frac{1}{c}\XM +\frac{c}{2}\XP ~.
\end{align}
Like before, the dilaton $X$ does not expand. The action \eqref{eq:mystrangeaddiction} then reduces to
\begin{align}
    \Bar{I}_{\textrm{\tiny Lorentz}}=&\frac{c\kappa}{2\pi}\int \Big[X\dd \omega +\XH \big(\dd \tau +\omega \wedge e\big)+\XP \dd e+\XM \big(\dd m +\omega \wedge \tau \big)+\XZ \dd z\nonumber\\
    &\qquad\quad +\Big(V(X)-\frac{U(X)}{2}\big(\XH^2-2\XM \XP \big)\Big)\,\tau \wedge e\Big]+\mathcal{O}(c^3)
\end{align}
which for $\kappa \to c^{-1}k$ and $c\to 0$ is the postcarrollian action \eqref{eq:action} with the potential \eqref{eq:UV_family}.


\section{All classical solutions}\label{se:4}

In this section, we derive all solutions to the classical equations of motion \eqref{eq:EOM_PSM} for the family of models \eqref{eq:UV_family}. These models automatically also include possible extensions of the potential \eqref{eq:UV_family} by $\XZ$-dependent shifts since $\XZ $ is a Casimir, $\dd \XZ=0$ on-shell. Such dependencies can therefore always be interpreted as a constant shift of the function $V(X)$. The postcarrollian CGHS model in section \ref{se:2.2} is such an example as it is given by $\mathcal{V}=\XZ=\text{cst}$. 

Another Casimir is $\XM$, which we shall use to classify solutions depending on whether it vanishes or not. In either of the cases, it is useful to define integrated versions of the potentials,
\begin{align}
Q(X):=\int^XU(y)\extd y\qquad\qquad w(X):=\int^Xe^{Q(y)}V(y)\extd y
\end{align}
in terms of which the last Casimir function is given by
\begin{align}\label{eq:nonlinear_C}
\mathcal{C}=w(X)-e^{Q(X)}\big(\tfrac12\,\XH^2-\XM\XP\big)\,.
\end{align}
On-shell it satisfies $\extd \mathcal{C}=0$ and we recast \eqref{eq:nonlinear_C} as
\eq{
\tfrac12\,\XH^2-\XM\XP = e^{-Q(X)}\,\big(w(X)-\mathcal{C}\big) := K(X)\,.
}{eq:Casimir_relation}

\subsection[Vanishing \texorpdfstring{$X_M$}{XM}]{Vanishing \texorpdfstring{$\boldsymbol{X_M}$}{XM}}\label{se:4.1}
We assume first $\XM=0$. In this case, the target space coordinate $\XH$ is Carroll boost invariant, and the classical solutions are quite similar to Carrollian black holes. Thus, we follow the procedure explained in \cite{Ecker:2023uwm} and obtain again two sectors. In the constant dilaton sector we have $\XH=0$ and constant $X$, subject to the constraint $V(X)=0$. For the remainder of this section, we focus on the linear dilaton sector, where $\XH\neq 0$ generically (though we could have $\XH=0$ at isolated loci, Carroll extremal surfaces).

Since \eqref{eq:torsion_eq} implies vanishing intrinsic torsion, $\dd e =0$, we choose our spatial coordinate $r$ such that $e=\dd r$ and find that \eqref{eq:carr_metric_eq} implies
\begin{align}\label{eq:X_eq}
    e=\extd r=-\frac{\extd X}{\XH}=\frac{\extd X}{\mp\sqrt{2K(X)}} 
\end{align}
where we fixed the Carroll boost freedom $\lambda_{\textrm{\tiny B}}$ by setting $\XP=0$. Using moreover the gauge freedom $\lambda _{\textrm{\tiny H}}$ to set $\tau =\tau _t\dd t$ we arrive at the solutions
\begin{align}
    e&=\dd r & \tau &=-e^{Q(X)}\,\XH\,\extd t & \omega&=e^{Q(X)}\big(V(X)-K(X)U(X)\big)\,\extd t\, \\
    \XP &=0 & \XH &=\pm \sqrt{2K(X)} & X(r)&:\,\text{by integrating \eqref{eq:X_eq}.}
\end{align}
The remaining one-form $m$ solves a trivial postcarrollian torsion condition,
\eq{
\extd m = 0
}{eq:copycat}
and $\XZ$ is any constant.  Thus, as far as the bulk geometry is concerned, the postcarrollian sector $\XM=0$ is indistinguishable from Carrollian solutions, and essentially all the results of \cite{Ecker:2023uwm} apply. For comparison with the sector $\XM\neq 0$, we stress that the dilaton field is here a spatial coordinate such that these solutions are interpreted as static postcarrollian spacetimes. 

\subsection[Non-vanishing \texorpdfstring{$X_M$}{XM}]{Non-vanishing \texorpdfstring{$\boldsymbol{X_M}$}{XM}}\label{se:4.2}

We assume here $\XM\neq 0$. Using the Carroll vector field $v^\mu $ defined by $v^\mu \tau _\mu =-1$, $v^\mu e_\mu =0$, the Carroll metric equation \eqref{eq:carr_metric_eq} implies $v^\mu \partial _\mu X\neq 0$. The dilaton describes a temporal direction in contrast to the $\XM=0$ sector. 

We define a new 1-form $Z$ by
\eq{
e = e^{Q(X)}\,Z
}{eq:e}
and use the Carroll metric equation \eqref{eq:carr_metric_eq} to obtain the clock 1-form
\eq{
\tau = -\frac{1}{\XM}\,\Big(\extd X + e^{Q(X)}\XH\,Z\Big) ~.
}{eq:tau}
The volume form is therefore given by
\eq{
\tau\wedge e = -\frac{e^{Q(X)}}{\XM}\,\extd X \wedge Z
}{eq:vol}
which reduces the intrinsic torsion equation to
\eq{
\extd Z=0\qquad\Rightarrow\qquad Z=\extd r ~.
}{eq:z}
Like before, we introduced a spatial coordinate $r$.

Since the value of $\XM$ is not very important (as long as it is non-zero), we pick $\XM=-1$, without loss of generality. We fix the Carroll boost freedom $\lambda _{\textrm{\tiny B}}$ by choosing $\XH=0$. As a consequence, the results above simplify to
\eq{
X:=t\qquad\qquad e=e^{Q(t)}\,\extd r\qquad\qquad\tau=\extd t\,.
}{eq:oida}
Thus, the dilaton field is the postcarrollian time in this solution sector. Moreover, there are no Carroll extremal surfaces. So this solution sector is qualitatively quite different from the vanishing $\XM$-sector.

Solving the postcarrollian Casimir equation for the boost connection yields
\eq{
\omega = e^{Q(t)}\big(K(t)U(t)-V(t)\big)\,\extd r\,.
}{eq:connection_sol}
The result \eqref{eq:connection_sol} is automatically compatible with the Carroll curvature and torsion conditions.

The remaining target space coordinate is obtained from the Casimir relation \eqref{eq:Casimir_relation},
\eq{
\XP = e^{-Q(t)}\,\big(w(t)-C\big)
}{eq:xp}
which leaves as the only remaining equation of motion the postcarrollian torsion equation \eqref{eq:postcalltorsioneq}. It simplifes to
\eq{
\extd m = \extd r\wedge\extd w(t)
}{eq:pc41}
and a possible solution of \eqref{eq:pc41} is
\eq{
m = -w(t)\,\extd r\,.
}{eq:m}

In summary, the gauge-fixed solutions for the sector $\XM\neq 0$ read
\begin{align}
    e&=e^{Q(X)}\dd r & \tau &=\dd t & \omega&=e^{Q(X)}\big(K(X)U(X)-V(X)\big)\,\extd r\, \\
    \XP &=K(X) & \XH &=0 & X&=t \\
    \XM &= -1 & m&=-w(X)\dd r ~. & &
\end{align}
The time dependence makes these solutions similar to cosmologies rather than black holes. 


\section{Boundary actions}\label{se:5}

In this section, we shall investigate a few possibilities of boundary conditions for 2d postcarrollian gravity theories, restricting to those that admit formulations as BF-theories for simplicity. This requires the Poisson tensor to be linear in the target space coordinates. After recalling some generalities from the literature, we shall restrict to the examples of the postcarrollian JT and CGHS models introduced in section \ref{se:2}. 

Since the spacetime manifold $\mathcal{M}$ is assumed to have a boundary $\partial \mathcal{M}$, the bulk action in general needs to be supplemented by a boundary action
\eq{
    \Gamma[\chi,\,A]=\frac{k}{2\pi}\,\int _{\mathcal{M}}\langle \chi ,F\rangle +I_{\partial \mathcal{M}} ~.
}{eq:full_BF_action}
This ensures that the variational principle is well-defined for a given set of boundary conditions of the dynamical variables, i.e., $\delta\Gamma$ vanishes on the solutions of the equations of motion, 
\eq{
    \delta \Gamma=(\text{equations of motion})+\frac{k}{2\pi}\int _{\partial \mathcal{M}}\langle \chi ,\delta A\rangle +\delta I_{\partial \mathcal{M}}\overset{!}{=}0 ~.
}{eq:IEOM}
The variation of $I_{\partial \mathcal{M}}$ therefore has to be fixed as
\eq{
    \delta I_{\partial \mathcal{M}}=-\frac{k}{2\pi}\int _{\partial \mathcal{M}}\langle \chi ,\delta A\rangle ~.
}{eq:var_bound}
Functionally integrating this expression to obtain $I_{\partial \mathcal{M}}$ requires imposing boundary conditions on $\chi $ and $A$ at $\partial \mathcal{M}$ \cite{Gonzalez:2018enk}. A general choice to integrate \eqref{eq:var_bound} is
\eq{
    \chi \big \vert _{\partial \mathcal{M}}\dd f= A\big \vert _{\partial \mathcal{M}}
}{eq:gen_BC}
where $f$ is some monotonous function on the boundary. Calling the boundary coordinate $t$, we choose $f=t$, although generalizations are possible. The boundary action integrates to 
\eq{
    I_{\partial \mathcal{M}}=-\frac{k}{4\pi}\int _{\partial \mathcal{M}}\dd t \,\langle \chi ,\chi \rangle 
}{eq:ilomilo}
where we implicitly assumed that $\partial \mathcal{M}$ is connected and either a line or a circle. This shows that, as usual for 2d dilaton gravity models \cite{Grumiller:2017qao,Gonzalez:2018enk}, the on-shell action is essentially given by the quadratic Casimir \eqref{eq:BF_Casimir} of the theory.

Since the boundary relation \eqref{eq:gen_BC} fixes the components of $\chi$ in terms of $A$ projected onto the boundary, we focus on the behavior of the latter and are assured that the variational principle is well defined by invoking \eqref{eq:gen_BC}. 

From the equation of motion obtained by varying $\chi$ in \eqref{eq:full_BF_action}, we find $F=0$ and therefore know that the field $A$ has to be pure gauge on-shell, $A=\Tilde{g}^{-1}\dd \Tilde{g}$. We choose a boundary condition for $A$ that is compatible with this solution,
\begin{align}\label{eq:A-bc}
    A&=b^{-1}\big(\dd +a\big)b & b=e^{rP}
\end{align}
where the boundary gauge field $a$ only has a leg in the $t$-direction,
\eq{
a=\big(\aB(t)B+\aH(t)H+\aP(t)P+\aM(t)M+\aZ(t)Z\big)\dd t 
}{eq:BG_Afield}
and all the dependence on the spatial coordinate $r$ is absorbed into the fixed group element $b$. We take $r$ as the direction normal to the boundary. A similar parametrization applies to the field $\chi$,
\eq{
    \chi =b^{-1}\,x\,b\qquad\qquad x=\xB(t)B+\xH(t)H+\xP(t)P+\xM(t)M+\xZ(t)Z ~.
}{eq:chibc}
The equations of motion reduce to 
\eq{
    \dd a+a\wedge a=0\qquad\qquad
    \dd x +[a,x] =0
}{eq:boundaryeom}
such that on-shell we have $a=g(t)^{-1}\dd g(t)$ with some $t$-dependent group element $g(t)$. 

In that case, the boundary condition \eqref{eq:gen_BC} reduces the action \eqref{eq:full_BF_action} to a pure boundary term
\eq{
    \Gamma[g]=-\frac{k}{4\pi}\int \dd t\, \langle g^{-1}g^\prime,\,g^{-1}g^\prime\rangle
}{eq:loop_action}
which is the action of a particle on the group manifold generated by $\mathfrak{g}$ \cite{Mertens:2017mtv,Mertens:2018fds,Gonzalez:2018enk,Grumiller:2020elf}. The integrand features the components of the Maurer--Cartan form $g^{-1}\dd g$, where we denote $\partial_t$ by prime. The equations of motion generated by this action are compatible with the bulk equations of motion \eqref{eq:boundaryeom}. The global symmetries of the action \eqref{eq:loop_action} are given by left- or right-multiplication with constant group elements,
\eq{
g\to g\,h \qquad\qquad g\to\tilde h\,g\qquad\qquad h,\,\tilde h = \textrm{cst}\,. 
}{eq:gs}
The left and right global symmetries \eqref{eq:gs} are not all independent from each other if some $g$, $h$, and $\tilde h$ commute, so the total number of global symmetries of the boundary action always lies between $N$ and $2N$, where $N$ is the rank of the gauge group. 

\subsection{Loop group boundary conditions for postcarrollian JT and CGHS}\label{se:5.1}

\newcommand{\para}{a}
\newcommand{\parb}{b}
\newcommand{\parc}{c}
\newcommand{\pard}{d}
\newcommand{\paralpha}{\alpha}
\newcommand{\parbeta}{\beta}
\newcommand{\pargamma}{\gamma}

\newcommand{\parepsb}{\epsB}
\newcommand{\parepsp}{\epsP}
\newcommand{\parepsh}{\epsH}
\newcommand{\parepsm}{\epsM}
\newcommand{\parepsz}{\epsZ}
\newcommand{\partepsb}{\Tilde{\varepsilon}_{\textrm{\tiny B}}}
\newcommand{\partepsp}{\Tilde{\varepsilon}_{\textrm{\tiny P}}}

We start with the postcarrollian JT model \eqref{eq:JT_action} with $\Lambda=-1$. Since the abelian 1-form $z$ is trivial for this model, we set it to zero from the start and also set to zero the corresponding target space coordinate $\XZ$. Then we are left with the Carroll boost connection $\omega$, the clock 1-form $\tau$, the spatial einbein $e$, the postcarrollian einbein $m$, and the associated target space coordinates $X,\XH,\XP,\XM$. 

It turns out to be convenient to change the basis, $L_\pm=H\pm B$, implying 
\eq{
[L_+,\,L_-] = 2M\qquad\qquad [L_\pm,\,P]=\pm L_\pm\qquad\qquad \langle L_+,\,L_-\rangle = -2\qquad\qquad\langle M,\,P\rangle = 1
}{eq:whatwasimadefor}
for the choice $\mu=1$ in \eqref{eq:wtf}. Assuming all components of the boundary connection 
\eq{
a=\big({\cal L}_+(t)\, L_+ + {\cal L}_-(t)\, L_- + {\cal P}(t)\, P + {\cal M}(t)\, M\big)\,\extd t
}{eq:CJT42}
are allowed to fluctuate, the boundary fields $a$, $x$ translate into the bulk variables by using \eqref{eq:A-bc} and \eqref{eq:chibc}, 
\begin{align}
A&=P\,\extd r+\big(e^r{\cal L}_+\,L_++e^{-r}{\cal L}_-\,L_-+{\cal P}\,P+{\cal M}\,M\big)\,\extd t\\
{\cal X}&=e^r x_+\,L_+ + e^{-r} x_-\,L_-+\xP\,P+\xM\,M\,.
\end{align}
They are preserved by the gauge transformations $\lambda=b^{-1}\varepsilon b$ with
\eq{
\varepsilon =\varepsilon_+(t)L_++\varepsilon_-(t)L_-+\epsP(t)P+\epsM(t)M 
}{eq:CJT43}
that act on the state-dependent functions as 
\begin{align}
\delta_\varepsilon {\cal L}_\pm &= \varepsilon^\prime_\pm \pm {\cal L}_\pm\,\epsP \mp {\cal P}\,\varepsilon_\pm \\
\delta_\varepsilon {\cal P} &= \epsP^\prime  \\
\delta_\varepsilon {\cal M} &= \epsM^\prime +2{\cal L}_+\,\varepsilon_- - 2{\cal L}_-\,\varepsilon_+\,.
\end{align}

Choosing the boundary group element
\eq{
g=e^{y\,L_+}e^{y_-\,L_-}e^{p\,P}e^{m\,M}
}{eq:CJT44}
one finds the Maurer--Cartan form by repeatedly using the Baker--Campbell--Hausdorff formula (see appendix \ref{app:A})
\eq{
g^{-1}g^\prime = y^\prime e^p\,L_++y_-^\prime e^{-p}\,L_-+p^\prime\,P+\big(m^\prime+2y^\prime y_-\big)\,M=a_t
}{eq:CJT45}
and can, as an alternative to \eqref{eq:CJT42}, express the field $A$ in terms of the fields $\{y,y_-,p,m\}$. The global symmetries \eqref{eq:gs} act on the functions in the boundary group element \eqref{eq:CJT44} as\footnote{%
After the Hamiltonian reduction in section \ref{se:5.2}, only the parameters $\para,\parb,\parc,\pard$ generate global symmetries of the reduced boundary action, while $\paralpha,\parbeta,\pargamma$ are no longer present. This motivates our notation. Also, we require $\parc\neq0$.
}
\begin{align}\label{eq:randomeqname}
    y &\to \parc\,y + \para + \paralpha\,e^{-p} & y_- &\to \frac{1}{\parc}\,y_--\frac{\parb}{2\parc} + \parbeta\,e^p \\
    p &\to p + \pargamma & m &\to m + \parb\,y+\pard-2\paralpha\,e^{-p}\,y_- 
\end{align}
which shows they are Goldstone-like fields as no derivatives of the transformation parameters appear. The boundary action \eqref{eq:loop_action} reads
\eq{
\Gamma[y,\,y_-,\,p,\,m] = \frac{k}{2\pi}\,\int\extd t\,\big(2y^\prime y_-^\prime-p^\prime(m^\prime+2y^\prime\,y_-)\big)
}{eq:CJT46}
and has the seven global symmetries \eqref{eq:randomeqname}. The general solution to the equations of motion descending from the boundary action \eqref{eq:CJT46},
\eq{
y=y_0+y_1\,e^{-p_1\,t}\qquad y_-=y_2+y_3\,e^{p_1\,t}\qquad p = p_0+p_1\,t\qquad m=m_0+m_1\,t-2y_1y_2\,e^{-p_1\,t}
}{eq:CJT48}
contains the eight integration constants $y_{0,1,2,3}$, $p_{0,1}$, and $m_{0,1}$. 

Let us turn to the postcarrollian CGHS model \eqref{eq:CGHS_action}. We pick a specific $\mathfrak{g}$ and a bilinear form on it given by \eqref{eq:CGHS_algebra}, \eqref{eq:CGHS_bil}. In this case the boundary fields $a$, $x$ translate into the bulk variables by using \eqref{eq:A-bc} and \eqref{eq:chibc}, $A_r=P$, and
\begin{align}\label{eq:randomname}
    A_t&= \aP (t)\,P+\aB(t)\,B + \big(r\,\aB(t)+\aH(t)\big)\,H + \aM(t)\,M + \Big(r\,\aH(t)+\aZ(t)+\frac{r^2}{2}\aB(t)\Big)\,Z \\
    {\cal X}&=\xB(t)\,B+(\xH(t)+\xB (t)r)\,H+\xP(t)\,P+\xM(t)\, M+\Big(\xZ(t)+\xH(t) r+\frac{r^2}{2}\xB(t) \Big)\,Z~.
\end{align}
The gauge transformations preserving \eqref{eq:BG_Afield} and \eqref{eq:chibc} are restricted to the form $\lambda =b^{-1}\varepsilon (t)b$ where
\begin{align}\label{eq:res_trafos}
    \varepsilon =\epsB(t)\,B+\epsH(t)\,H+\epsP(t)\,P+\epsM(t)\,M+\epsZ(t)\,Z ~.
\end{align}
This gives five free functions at the boundary. These boundary fields infinitesimally transform as
\begin{align}
    \delta _\varepsilon \aB &=\epsB' & \delta _\varepsilon \xB&=0\\
    \delta _\varepsilon \aH&=\epsH'+\aB\epsP-\aP\epsB  & \delta _\varepsilon \xH&=\xB\epsP-\xP\epsB\\
      \delta _\varepsilon \aP&=\epsP'  & \delta _\varepsilon \xP&=0\\
    \delta _\varepsilon \aM&=\epsM'+\aB\epsH-\aH\epsB & \delta _\varepsilon \xM&=\xB\epsH-\xH\epsB\\
    \delta _\varepsilon \aZ&=\epsZ'+\aH\epsP-\aP\epsH  & \delta _\varepsilon \xZ&=\xH\epsP-\xP\epsH ~.
\end{align}

Choosing the boundary group element as
\eq{
g(t)=e^{h(t)H}e^{p(t)P}e^{b(t)B}e^{m(t)M}e^{z(t)Z}
}{eq:youshouldseemeinacrown}
one finds the Maurer--Cartan form by repeatedly using the Baker--Campbell--Hausdorff formula (see appendix \ref{app:A})
\eq{
    g^{-1}g^\prime=b^\prime B+(h^\prime-p^\prime b)H+p^\prime P+\Big(m^\prime-h^\prime b+\frac{1}{2}p^\prime b^2\Big)M +(z^\prime+h^\prime p)Z =a_t
}{eq:badguy}
and can, as an alternative to \eqref{eq:BG_Afield}, express the field $a$ in terms of the fields $\{b,h,p,m,z\}$. The global symmetries \eqref{eq:gs} act on these as 
\begin{subequations}\label{eq:goldstone_trafo}
\begin{align}
    b&\to b+\parepsb \\
    h&\to h+\parepsh+b\,\parepsp+p\parepsb \\
    p&\to p+\parepsp \\
    m&\to m+\parepsm +b\,\parepsh +\frac{b^2}{2}\,\parepsp + \frac{p}{2}\,\partepsb^2 \\
    z&\to z+\parepsz -p\,\parepsh -bp\,\parepsp-\frac{b}{2}\,\parepsp^2 - h\,\partepsp - p\,\partepsb\partepsp-\frac{p^2}{2}\,\partepsb 
\end{align}
\end{subequations}
which shows they are Goldstone-like fields as no derivatives of the gauge parameters appear. The boundary action \eqref{eq:loop_action} reads
\eq{ 
    \Gamma[b,\,h,\,p,\,m,\,z]=\frac{k}{2\pi}\int \dd t \, \Big(p^\prime m^\prime+b^\prime\,\big(z^\prime+ph^\prime \big) -\frac12\,h^{\prime\,2}\Big) 
}{eq:lg}
and has the global symmetries \eqref{eq:goldstone_trafo}. The field equations descending from the loop group boundary action \eqref{eq:lg} yield monomial expressions in time depending on ten integration constants,
\begin{subequations}
    \label{eq:whatever}
\begin{align}
    b&=b_0+b_1t & m&=m_0+m_1t+\frac{b_1h_1}{2}t^2-\frac{b_1^2p_1}{6}t^3  \qquad\qquad h=h_0+h_1t-\frac{b_1p_1}{2}t^2 \\
    p&=p_0+p_1t &z&=z_0+z_1t+\frac{p_1(p_0b_1-h_1)}{2}t^2+\frac{p_1^2b_1}{3}t^3\,. 
\end{align}
\end{subequations}

The loop group boundary actions \eqref{eq:CJT46} and \eqref{eq:lg} are the postcarrollian analogs of the WZW-action descending from Chern--Simons theories \cite{Witten:1989hf,Elitzur:1989nr}. Typically, one is interested in stricter boundary conditions that constrain the boundary dynamics in a geometrically interesting way. A famous example is the Drinfeld--Sokolov reduction of sl$(2,\mathbb{R})$ current algebras to Virasoro algebras, which yields Brown--Henneaux boundary conditions \cite{Brown:1986nw} when applied to AdS$_3$ Einstein gravity \cite{Coussaert:1995zp}. 

For the case of JT gravity with a negative cosmological constant, this, for example, leads to the Schwarzian action \cite{Maldacena:2016hyu,Sarosi:2017ykf,Mertens:2018fds,Gu:2019jub} as a reduction of the AdS$_2$-loop group boundary action \cite{Gomis:2020wxp,Grumiller:2020elf}. While there are several methods to perform this reduction, the one we employ here is the inverse Higgs mechanism \cite{Gomis:2020wxp,Galajinsky:2019lak}. It amounts to fixing certain components of the Maurer--Cartan form and solving the resulting constraints for the remaining fields in the boundary action.

\subsection{Schwarzian-type boundary actions for postcarrollian JT and CGHS}\label{se:5.2}

The postcarrollian JT algebra \eqref{eq:angelinajolie} has an alternative interpretation as centrally extended AdS--Carroll algebra (which in turn is isomorphic to the centrally extended Poincar\'e algebra) studied already in \cite{Grumiller:2020elf,Gomis:2020wxp}. Therefore, we simply copy the results from section 6.2 of \cite{Grumiller:2020elf} to get a Schwarzian-type boundary action. Below, we briefly summarize, simplify, and expand these results. As before, we set $\Lambda=-1$ and $\mu=1$ in \eqref{eq:wtf}.

After the same basis change, $L_\pm = H\pm B$ [see \eqref{eq:whatwasimadefor}], we fix boundary conditions in highest-weight gauge with the ansatz \eqref{eq:A-bc} and the boundary connection
\eq{
a=\big(L_+-{\cal L}(t)\,L_-+{\cal T}(t)\,P\big)\,\extd t\,. 
}{eq:CJT1}
The gauge parameters preserving these boundary conditions, $\lambda=b^{-1}\varepsilon b$ with $\varepsilon=\varepsilon_+\,L_++\varepsilon_-\,L_-+\varepsilon_{\textrm{\tiny P}}\,P+\varepsilon_{\textrm{\tiny M}}\,M$, generate the conditions $\varepsilon_{\textrm{\tiny P}}={\cal T}\,\varepsilon_+-\varepsilon_+^\prime$, $\varepsilon_-=-{\cal L}\varepsilon_+-\frac12\varepsilon_{\textrm{\tiny M}}^\prime$, and the transformations
\eq{
\delta_\varepsilon{\cal T}=\lambda_+{\cal T}^\prime+\lambda_+^\prime{\cal T}-\lambda_+''\qquad\qquad\delta_\varepsilon {\cal L}=\lambda_+{\cal L}^\prime+2\lambda_+^\prime{\cal L}+\frac12\,\big(\lambda_{\textrm{\tiny M}}^\prime{\cal T}+\lambda_{\textrm{\tiny M}}''\big)
}{eq:CJT2}
on the state-dependent functions that are recognized as twisted warped transformations \cite{Afshar:2015wjm,Afshar:2019tvp,Afshar:2019axx,Afshar:2021qvi}.

In the second-order formulation, the boundary conditions \eqref{eq:CJT1} translate into the Carrollian metric and vector field
\eq{
h_{\mu\nu}\,\extd x^\mu\extd x^\nu = \big(\extd r+{\cal T}\,\extd t\big)^2\qquad\qquad v^\mu\partial_\mu = -\frac{1}{e^r-{\cal L}\,e^{-r}}\,\big(\partial_t-{\cal T}\,\partial_r\big)
}{eq:CJT3}
which straightforwardly generalize to the Brown--Henneaux type conditions
\eq{
h_{\mu\nu}\,\extd x^\mu\extd x^\nu = \big(\extd r + {\cal O}(1)\,\extd t\big)^2\,\big(1+{\cal O}(e^{-2r})\big)\qquad v^\mu\partial_\mu = -e^{-r}\,\big(\partial_t+{\cal O}(1)\partial_r\big)\,\big(1+{\cal O}(e^{-2r})\big)
}{eq:CJT3a}
subject to the orthogonality constraint $h_{\mu\nu}v^\nu=0$ to all orders. Applying the analysis of \cite{Ecker:2023uwm} to zero mode solutions (constant ${\cal T,L}$), we deduce there is a Carroll extremal surface at $e^{2r}={\cal L}$. 

Parametrizing the postcarrollian JT group element as
\eq{
g = e^{y\,L_+}e^{y_-\,L_-}e^{p\,P}e^{m\,M}
}{eq:CJ4}
the boundary conditions \eqref{eq:CJT1} impose the constraints
\eq{
g^{-1}g^\prime\big|_{L_+} = 1\qquad\qquad g^{-1}g^\prime\big|_{M} = 0 
}{eq:CJT5}
and yield the Schwarzian-type boundary action \cite{Grumiller:2020elf,Gomis:2020wxp}
\eq{
\Gamma[y,\,m] = \frac{k}{2\pi}\,\int\extd t\,\big(
m^\prime\ln^\prime y^\prime-m'' 
\big)\,. 
}{eq:CJT6} 
The boundary action \eqref{eq:CJT6} is invariant under the global postcarrollian transformations
\eq{
y\to \parc\,y + \para\qquad\qquad m \to m + \parb\,y + \pard\,.
}{eq:CJT7}

The state-dependent functions 
\eq{
{\cal T}=-\ln^\prime y^\prime \qquad\qquad {\cal L}=-\frac12\,\big(m^\prime\ln^\prime y^\prime-m''\big) 
}{eq:CJT8}
are conserved on-shell, $\partial_t{\cal T}=0=\partial_t{\cal L}$, which is shown by varying the twisted warped boundary action \eqref{eq:CJT6} with respect to the time parametrization field $y$ and the phase field $m$. 

Let us turn to the postcarrollian CGHS model. One possibility for a set of constraints on the boundary variables is to fix
\begin{align}
    g^{-1}g^\prime\big\vert_B &=1 &  g^{-1}g^\prime\big\vert_Z&=0
\end{align}
solved by
\begin{align}
    b&=b_0+t  & z^\prime &=-h^\prime p  ~.
\end{align}
Defining the time- and state-dependent functions
\eq{
\mathcal{T}=p^\prime\qquad\qquad\mathcal{P}=tp^\prime-h^\prime\qquad\qquad\mathcal{M}=m^\prime-t\,h^\prime+\frac{t^2}{2}\,p^\prime
}{eq:buryafriend}
the Carrollian geometry is given by
\eq{
h_{\mu\nu}\,\extd x^\mu\extd x^\nu = \big(\extd r + \mathcal{T}\,\extd t\big)^2\qquad\qquad v^\mu\partial_\mu = \frac{1}{r-\mathcal{P}}\,\big(\partial_t-\mathcal{T}\,\partial_r\big)
}{eq:therforeiam}
and for zero mode solutions features a Carroll extremal surface at $r=\mathcal{P}$. The postcarrollian 1-form 
\eq{
m=\mathcal{M}\,\extd t
}{eq:idontwannabeyouanymore}
is independent of the spatial coordinate $r$ and only has a leg in the time direction.

The loop group boundary action \eqref{eq:lg} reduces to 
\eq{
    \Gamma[h,\,p,\,m]=\frac{k}{2\pi}\int \dd t \,\Big(p^\prime m^\prime-\frac12\,h^{\prime\,2}\Big)
}{eq:birdsofafeather}
which resembles three free particles moving along the boundary, since on-shell one finds $h''=p''=m''=0$. Equivalently, the equations of motion are generalized conservation equations,
\eq{
\partial_t\mathcal{T}=0\qquad\qquad\partial_t\mathcal{P}=\mathcal{T}\qquad\qquad\partial_t\mathcal{M}=\mathcal{P}-t\,\mathcal{T}\quad\Rightarrow\quad\partial_t^2\mathcal{P}=\partial_t^2\mathcal{M}=0\,.
}{eq:whentehpartysover}

Imposing the constraints on the level of the transformation laws \eqref{eq:goldstone_trafo}, one finds a reduced set of symmetries for the remaining fields
\eq{
h\to h + \parepsh\qquad\qquad p\to p+\parepsp\qquad\qquad m\to m+\parepsm  \,. 
}{eq:happierthanever}

While these boundary conditions lead to quite simple dynamics in these variables, it might still be possible to impose different constraints such that one does not find just free particles. However, such a choice of constraints would have to be more sophisticated than just imposing some of the Maurer--Cartan components to vanish, since all such combinations we investigated lead to similar conclusions. 

As an alternative to the inverse Higgs mechanism, one can apply the technique presented in \cite{Grumiller:2020elf}. This amounts to constructing constraints as a consistent second-class system on the level of the algebra. The possible choices depend on the number of even-dimensional non-abelian subalgebras $\mathfrak{h}$ of $\mathfrak{g}$ which have $\langle \mathfrak{h},\,K\rangle \neq 0$ for some fixed algebra element $K\notin \mathfrak{h}$. For \eqref{eq:CGHS_algebra} the only 4d choices of $\mathfrak{h}$ are $\{B,H,M,Z\}$ and $\{P,H,M,Z\}$, both of which again lead to free particle actions. There are no 2d choices. However, this conclusion relies on the fixed element $K\in \mathfrak{g}$ being chosen in a state-independent manner, which at the same time suggests the direction in which this simple result could be circumvented. 


\section{Generalization to higher dimensions}\label{se:6}

In this section, we generalize postcarrollian gravity to higher dimensions. We start with 3d in section \ref{se:6.1} and conclude with 4d in section \ref{se:6.2}.

\subsection{Postcarrollian gravity in 3d}\label{se:6.1}

In $D=2+1$ dimensions, it is possible to define a postcarrollian gravity theory by considering a Chern--Simons action
\begin{equation}\label{CSaction}
I_{3d}=\frac{k}{4\pi}\int \left\langle A \wedge dA + \frac23\, A\wedge A\wedge A \right\rangle
\end{equation}
where the connection one-form $A=A_\mu \extd x^\mu$ follows from \eqref{connectionA}, and $\left\langle\,,\,\right\rangle$
denotes a non-degenerate invariant bilinear form on the postcarrollian algebra \eqref{PostCalgebra}. By defining the dual generators
\begin{align}\label{dualgen3d}
    J&=\frac{1}{2}\epsilon ^{ab}J_{ab} & S&=\frac{1}{2}\epsilon ^{ab}S_{ab} & C_a&=-\epsilon _a{}^b B_b & G_a&=-\epsilon _a{}^bK_b~.
\end{align}
where $\epsilon_{ab}$ is the 2d spatial Levi-Civita symbol ($\epsilon_{12}=1$), the commutation relations \eqref{PostCalgebra} yields
\begin{multicols}{2}
\begin{subequations}
\label{PostCalgebra3D}
\setlength{\abovedisplayskip}{-14pt}
\allowdisplaybreaks
\begin{align}
    [J,P_a]&=\epsilon^b{}_a\,P_b \\
    [J,C_a]&=\epsilon^b{}_a\,C_b\\
    [P_a,P_b]&=-\Lambda\, \epsilon_{ab}\,J \\
    [P_a,C_b]&=-\epsilon_{ab}\,H \\
    [P_a,H]&=-\Lambda\, \epsilon^b{}_a\,C_b \\
    [C_a,C_b]&=\epsilon_{ab}\,S \\
    [C_a,H]&=\epsilon^b{}_a\,M_b \\
    [J,M_a]&=\epsilon^b{}_a\,M_b=[S,P_a] \\
    [J,G_a]&=\epsilon^b{}_a\,G_b=[S,C_a] \\
    [P_a,M_b]&=-\Lambda\, \epsilon_{ab}\,S \\
    [P_a,G_b]&=-\epsilon_{ab}\,L= [M_a,C_b] \\
    [P_a,L]&=-\Lambda\,\epsilon^b{}_a\,G_b=[M_a,H] \\\nonumber\\&\!\!\!\!\!\!\!\!\!\!\!\!\!\!\!\! \textrm{commutators\;not\;displayed\;vanish.} \nonumber
\end{align}
\end{subequations}
\end{multicols}
\noindent Up to dualization with the Levi-Civita symbol, the algebra above is the specialization of the general postcarrollian algebra \eqref{PostCalgebra} to 3d. Again, the commutators in the left half form the Carroll algebra provided we set to zero the right-hand sides in the last two lines. The commutators in the right half all involve one subleading generator on each side of the equalities that mirror the corresponding leading generators from the left half.

In $2+1$ dimensions, the postcarrollian algebra admits the invariant bilinear form 
\begin{subequations}\label{IBFPostCalgebra}
\begin{align}
    \langle J,H \rangle &=\rho_0 & \langle S,H \rangle &=\rho_1 = \langle J,L \rangle \\
    \langle P_a,C_b \rangle &=\rho_0\,\delta_{ab} & 
    \langle P_a,G_b \rangle &=\rho_1\,\delta_{ab} = \langle M_a,C_b \rangle 
\end{align}
\end{subequations}
where $\rho _0$, $\rho _1\in \mathbb{R}$ and $\rho _1\neq 0$ for $\langle \cdot , \cdot \rangle $ to be non-degenerate. By defining dual gauge fields compatible with \eqref{dualgen3d}, the postcarrollian connection one-form is written as\footnote{%
The generators for subleading boosts and rotations $G_a$ and $S$ are the dualized versions, but we still use the same letter for the corresponding gauge field components.}
\begin{align}
    A=\tau H+e^aP_a+\omega ^aC_a+\omega J+lL+m^aM_a+\sigma^aG_a+\sigma S ~.
\end{align}
From $\delta _\lambda A=\dd \lambda +[A,\lambda]$ and 
\eq{
    \lambda =\xi H +\xi ^a P_a+\theta ^aC_a+\theta J+\zeta L+\zeta ^aM_a+\eta ^aG_a+\eta S
}{eq:bellyache}
one can infer the gauge-transformation behaviour of the various components
\begin{subequations}
\label{eq:hostage}
\begin{align}
    \delta _\lambda \tau &=\dd \xi +\epsilon _{ab}(\xi ^a\omega ^b-e^a\theta ^b)\\
    \delta _\lambda e^a&=\dd \xi ^a+\epsilon ^a{}_b(\omega \xi ^b-\theta e^b)\\
    \delta _\lambda \omega &=\dd \theta -\Lambda e^a\epsilon_{ab}\xi ^b\\
    \delta _\lambda \omega ^a&=\dd \theta ^a+\epsilon ^a{}_b(\omega \theta ^b-\theta \omega ^b+\Lambda \tau \xi ^b-\Lambda \xi e^b)\\
    \delta _\lambda l&=\dd \zeta -\epsilon _{ab}(e^a\eta ^b-\xi ^a\sigma ^b+m^a\theta ^b-\zeta ^a\omega ^b)\\
    \delta _\lambda m^a&=\dd \zeta ^a+\epsilon ^a{}_b(\omega \zeta ^b-\theta m^b+\sigma \xi ^b-\eta e^b-\tau \theta ^b+\xi \omega ^b)\\
    \delta _\lambda \sigma &=\dd \eta +\epsilon _{ab}(\omega ^a\theta ^b-\Lambda e^a\zeta ^b+\Lambda \xi ^am^b)\\
    \delta _\lambda \sigma ^a&=\dd \eta ^a+\epsilon ^a{}_b(\omega \eta ^b-\theta \sigma ^b+\sigma \theta ^b-\eta \omega ^b+\Lambda l\xi ^b-\Lambda \zeta e^b+\Lambda \tau \zeta ^b-\Lambda \xi m^b)~.
\end{align}
\end{subequations}
The curvature two-form $F=\dd A + A\wedge A$ is given by 
\begin{multline}
F=T H+ T^a P_a
+\left(R^a(C)+\Lambda \epsilon ^a{}_b \tau \wedge e^b\right)C_a
+\left(R(J)-\frac{\Lambda}{2}\epsilon _{ab}e^a\wedge e^b\right)J
\\
+\Theta L+ \Theta^a M_a
+\left(R^a(G)+\Lambda \epsilon ^a{}_b\tau \wedge m^b+ \Lambda \epsilon ^a{}_b l\wedge e^b\right)G_a
+\left(R(S)-\Lambda \epsilon _{ab}e^a\wedge m^b\right)S
\end{multline}
where we have defined the Carrollian and postcarrollian torsion two-forms
\begin{subequations}
\begin{align}
   T&=\dd \tau -\epsilon _{ab}\,\omega ^a\wedge e^b \\
    T^a&=\dd e^a+\epsilon ^a{}_b\,\omega \wedge e^b \\
   \Theta &=\dd l-\epsilon _{ab}\left(\sigma^a\wedge e^b-\omega ^a\wedge m^b\right)\\
    \Theta^a &=\dd m^a+\epsilon ^a{}_b\left(\omega \wedge m^b+\sigma \wedge e^b-\tau \wedge \omega ^b\right)
\end{align}
\end{subequations}
and curvature two-forms
\begin{subequations}
\begin{align}
    R^a(C)&=\dd \omega ^a+\epsilon^a{}_b\,\omega \wedge \omega ^b\\
    R(J)&=\dd \omega \\
    R^a(G)&=\dd \sigma^a+\epsilon ^a{}_b\,\omega \wedge \sigma^b+\epsilon ^a{}_b\,\sigma \wedge \omega ^b\\
    R(S)&=\dd \sigma +\frac{1}{2}\epsilon _{ab}\,\omega ^a\wedge \omega ^b ~.
\end{align}
\end{subequations}

Using these expressions, the Chern-Simons action \eqref{CSaction} takes the form
\begin{equation}
    I_{\textrm{\tiny 3d}}=\rho_0 \, I_{\textrm{\tiny{(A)dSC}}}+ \rho_1\, I_{\textrm{\tiny PC}}
\end{equation}
where the first term corresponds to the magnetic (A)dS$_3$ Carroll gravity action \cite{Matulich:2019cdo,Ravera:2019ize,Concha:2024tcu}
\begin{equation}\label{eq:ADSCarroll-3d}
I_{\textrm{\tiny{(A)dSC}}}= \frac{k}{2\pi}\int \Big(e_a\wedge R^a(C)+\tau \wedge R(J)-\frac{\Lambda}{2} \epsilon _{ab}\tau \wedge e^a\wedge e^b\Big)
\end{equation}
whereas the second is given by the following postcarrollian gravity action.
\begin{equation}\label{PCaction3d}
I_{\textrm{\tiny PC}}= \frac{k}{2\pi}\int \Big(m_a\wedge R^a(C)+\tau \wedge R(S)+l\wedge R(J)+e_a\wedge R^a(G)-\frac{\Lambda}{2}\epsilon _{ab}\big(2\tau \wedge m^a\wedge e^b+l\wedge e^a\wedge e^b\big)\Big) 
\end{equation}
The field equations that follow from varying the (A)dS Carroll action \eqref{eq:ADSCarroll-3d} read
\begin{equation}\label{eq:3d_LO_equations}
T=0=T^a \qquad R^a (C) =-\Lambda \epsilon^a{}_b \tau \wedge e^b \qquad R(J) = \frac{\Lambda}2 \epsilon_{ab} e^a\wedge e^b
\end{equation}
The variations of the postcarrollian action \eqref{PCaction3d} with respect to the postcarrollian fields reproduce field equations associated with the (A)dS Carroll action \eqref{eq:ADSCarroll-3d}. Varying the action with respect to Carrollian gauge fields yields a new set of field equations,
\begin{equation}\label{eq:eomPC}
\Theta=0=\Theta^a \qquad R^a (G) =-\Lambda \epsilon^a{}_b \left(\tau \wedge m^b+l\wedge e^b\right) \qquad R(S) = \Lambda \epsilon_{ab} e^a\wedge m^b ~.
\end{equation}
This nested structure of the equations of motion is typical for theories arising from algebra expansions and, e.g., also appears in non-relativistic expansions of general relativity \cite{Hansen:2020pqs}.

\subsubsection{Reduction from 3d to 2d}
Among the solutions of the 3d postcarrollian theory \eqref{PCaction3d} we just presented are the spherically symmetric ones, which admit an effective description in terms of the postcarrollian JT model we described in section \ref{se:2.1}. This is in direct analogy to the Lorentzian case \cite{Achucarro:1993fd,Mertens:2018fds}. Let us see which conditions spherical symmetry implies on the frame variables. For this, we use adapted coordinates $(x^\alpha,\phi)$ where we denote the 2d indices by $\alpha,\beta =0,1$. We also split 3d frame components into constituents adapted to the symmetry like $e^a=(e,e^{\2})$, $m^a=(m,m^{\2})$ and denote the values of anholonomic indices by $a,b=\1,\2$ with $\0$ denoting the component in the direction of $\tau $. In the ansatz for the spherical reduction,
\begin{align}
    e&=e_\alpha \dd x^\alpha & e^{\2}&=Y \dd \phi \\
    m&=m_\alpha \dd x^\alpha & m^{\2}&=X \dd \phi \\
    \tau &=\tau _\alpha \dd x^\alpha & l&=l_\alpha \dd x^\alpha 
\end{align}
all the components are independent of the coordinate direction $\phi$. It follows that some of the equations of motion reduce to kinematical constraints and only a part of them carries dynamical information. This dynamical part should be recovered as equations of motion of the reduced 2d theory. The kinematical part will be solved directly and inserted into the action.

As a first step, we therefore expand the torsion constraints
\eq{
    T^a=T=\Theta ^a=\Theta=0 
}{eq:3d_torsionconstraints}
in terms of their frame components like
\begin{align}
    T^a=T^a_{\0\1}\tau \wedge e +T^a_{\0\2}\tau \wedge e^{\2}+T^a_{\1\2}e\wedge e^{\2}
\end{align}
and similarly for the other torsion two-forms. The connection and vielbein components are likewise expanded as $\omega ^a=\tau \omega _{\0}^a+e\omega _{\1}^a+e^{\2}\omega _{\2}^a$ and $\sigma^a=\tau \sigma _{\0}^a+e\sigma _{\1}^a+e^{\2}\sigma ^a_{\2}$.  
The kinematical part of the first half in \eqref{eq:3d_torsionconstraints} implies
\begin{align}\label{eq:Tconstr1}
    \omega _{\0}&=0 & \omega _{\1}&=0 & \omega ^{\1}_{\1}&=-\omega ^{\2}_{\2} & \omega ^{\1}_{\0}&=0
\end{align}
while the second half yields the constraints
\begin{multline}
    \omega _{\0}m^{\2}_{\2}+\sigma _{\0}-\omega ^{\2}_{\2} = 
    \omega _{\1}m^{\2}_{\2}+\sigma _{\1} = 
    \omega _{\0}m_{\1}-\omega _{\1}m_{\0}+\sigma _{\0}-\omega ^{\1}_{\1} \\
    = 
    \sigma ^{\1}_{\1}+\sigma ^{\2}_{\2}+\omega ^{\1}_{\1}m^{\2}_{\2}+\omega ^{\2}_{\2}m_{\1} = 
    \sigma ^{\1}_{\0}+\omega ^{\1}_0m^{\2}_{\2}-\omega ^{\2}_{\2}m_{\1}= 0\,.
\label{eq:NLO_constr}
\end{multline}
If used together with \eqref{eq:Tconstr1} these equations imply
\begin{align}
    \sigma^{\0}&=0 & \sigma^{\1}&=0 & \sigma^{\1}_{\1}&=-\sigma^{\2}_{\2} & \sigma^{\1}_{\0}&=0  & \omega ^{\1}_{\1}&=0=\omega ^{\2}_{\2}~.
\end{align}

The rest of the equations of motion \eqref{eq:ADSCarroll-3d}-\eqref{eq:eomPC} do not give any additional constraints but are either trivially satisfied or carry dynamical information. As usual for a Carrollian connection the torsion constraints in the first half of \eqref{eq:3d_torsionconstraints} do not determine $\omega ^a$ uniquely but only up to the shift $\omega ^a \to \omega  ^a+\epsilon ^{ab}W_{bc}e^c$ where $W_{ab}$ is an arbitrary symmetric tensor \cite{Bergshoeff:2017btm,Campoleoni:2022ebj}. The torsion constraints in the second half of \eqref{eq:3d_torsionconstraints} have a similar structure, $\sigma^a$ is only determined up to $\sigma^a \to \sigma^a +\epsilon ^{ab}V_{bc}e^c$.
Before looking at the full postcarrollian theory, let us see how things work in the (A)dS Carroll theory as a warm-up.

We solve the constraints \eqref{eq:Tconstr1},
\begin{align}
    \omega =\omega _\phi \dd \phi && \omega ^{\1}=-\omega ^{\2}_{\2} e+\omega ^{\1}_\phi \dd \phi && \omega ^{\2}=\Omega +\omega ^{\2}_{\2} X \dd \phi 
\end{align}
where $\Omega$ is some one-form with legs only in the $\tau$ and $e$ directions. The ambiguity $W_{ab}$ acts on these components by
\begin{align}
    \omega ^{\2}_{\2}\to \omega ^{\2}_{\2}-W_{{\1}{\2}} && \omega ^{\1}_\phi \to \omega ^{\1}_\phi +W_{{\2}{\2}} && \Omega \to \Omega -W_{{\1}{\1}} e~,
\end{align}
which allows to parametrize them as
\begin{align}
    \omega =\omega _\phi \dd \phi && \omega ^{\1}=W_{{\1}{\2}} e+W_{{\2}{\2}} \dd \phi && \omega ^{\2}=\Omega -W_{{\1}{\2}} X \dd \phi ~.
\end{align}
The action \eqref{eq:ADSCarroll-3d} then reads
\begin{align}
    I_{\textrm{\tiny (A)dSC}}=\frac{k}{2\pi} \int \dd \phi \wedge \Big[Y \dd \Omega +\omega _\phi \big(\dd \tau +\Omega \wedge e\big)+W_{{\2}{\2}} \dd e-Y \Lambda \tau \wedge e]
\end{align}
which after integrating over $\phi$ becomes the action of the Carrollian JT model \cite{Grumiller:2020elf} upon identifying
\begin{align}
    \XH =\omega _\phi && \XP =W_{{\2}{\2}} && \Omega = \omega  ~.
\end{align}
Note that $W_{{\1}{\2}}$ does not feature in the action. Moreover, the component $W_{{\1}{\1}}$ can always be absorbed into $\Omega$ and therefore does not feature in the action either.

The postcarrollian theory is obtained by solving all the kinematical equations, i.e., solving additionally \eqref{eq:NLO_constr}. This yields the general solution
\begin{align}
    \sigma &=\sigma _\phi \dd \phi & \sigma^{\1}&=V_{{\1}{\2}}e+V_{{\2}{\2}}\dd\phi & \sigma^{\2}&=B-V_{{\1}{\2}}Y\dd \phi \\
    \omega &=\omega _\phi \dd \phi & \omega ^{\1}&=W_{{\2}{\2}}\dd \phi & \omega ^{\2}&=\Omega 
\end{align}
where we introduced two one-forms $B=B_\alpha \dd x^\alpha$ and $\Omega =\Omega _\alpha \dd x^\alpha$ and all the ambiguous components were already parametrized by $V_{ab}$ and $W_{ab}$. 

Inserting this into the postcarrollian action \eqref{PCaction3d} we get up to total derivatives
\begin{align}
    I_{\textrm{\tiny PC}}= \frac{k}{2\pi} \int \dd \phi \wedge \Big[&X\dd \Omega +\sigma_\phi \big(\dd \tau +\Omega \wedge e\big)+V_{{\2}{\2}}\dd e\nonumber
    +W_{{\2}{\2}}\big(\dd m+\Omega \wedge \tau \big)-\Lambda X\tau \wedge e  \nonumber\\
    &+\omega _\phi \big(\dd l+B\wedge e\big)+Y\big(\dd B-\Lambda (\tau \wedge m +l\wedge e)\big)\Big]\,.
\end{align}
One can see that the 2d theory from previous sections appears as a subsector by solving the constraints associated with $\omega_\phi$ and $Y$, effectively setting the last line to zero. If we then identify 
\begin{align}
    \sigma _\phi =\XH && V_{{\2}{\2}}=\XP && W_{{\2}{\2}}=\XM && \Omega &=\omega 
\end{align}
we obtain the postcarrollian JT model \eqref{eq:JT_action} after integrating out the angular part. 

\subsubsection{Lifting Carrollian boundary conditions to postcarrollian ones}\label{se:6.1.3}

In 3d gravity theories, there is a menagerie of boundary conditions for a given bulk theory, see \cite{Grumiller:2016pqb,Grumiller:2017sjh} and Refs.~therein for examples within Einstein gravity. It is not the intention of this work to generalize such a discussion to postcarrollian gravity theories. Instead, we confine ourselves to one example to show that the same methods that worked for 3d Einstein gravity and 3d Carroll gravity continue to work for 3d postcarrollian gravity.

Our example is based on the Carrollian boundary conditions proposed in \cite{Bergshoeff:2016soe}. We adopt their ansatz and extend it to the postcarrollian gravity theory introduced at the beginning of section \ref{se:6.1}, fixing $\rho_0=0$ and $\rho_1=1$ in the bilinear form \eqref{IBFPostCalgebra} and choosing the value $\Lambda=0$ for the cosmological constant. Thus, we start with the bulk Chern--Simons connection
\eq{
A=b^{-1}(r)\,\big(\extd\,+\,a(t,\,\varphi)\big)\,b(r)\qquad\qquad b(r) = e^{r\,P_2}
}{eq:3d1}
where the boundary connection 
\eq{
a(t,\,\varphi)=\big(J+\aH\,H+\aP^b\,P_b+\aC^b\,C_b+\aL\,L+\aM^b\,M_b+\aG^b\,G_b+\aS\,S\big)\,\extd\varphi + \mu\,H\,\extd t
}{eq:3d2}
contains all the Carrollian and postcarrollian generators introduced above. The quantities with letters $a$ are all state-dependent functions that can depend on the postcarrollian time $t$ and the angular coordinate $\varphi\sim\varphi+2\pi$ but not on the radial coordinate $r$. The quantity $\mu$ is a chemical potential, i.e., an arbitrary state-independent function of $t,\varphi$.

The choices \eqref{eq:3d1}, \eqref{eq:3d2} yield the bulk connection 
\eq{
A=a+P_2\,\extd r+r\,[a,\,P_2]
}{eq:3d3}
that is linear in the radial coordinate $r$. The nonvanishing commutators with the translation generator $P_2$ are given by
\eq{
[J,\,P_2]=P_1\qquad\qquad[P_2,\,C_1]=H\qquad\qquad[S,\,P_2]=M_1\qquad\qquad[P_2,\,G_1]=L\,.
}{eq:3d4}
The metric deduced from \eqref{eq:3d3},
\eq{
h_{\mu\nu}\,\extd x^\mu\extd x^\nu=\extd r^2 + 2\aP^2\,\extd\varphi\extd r + \big((r+\aP^1)^2+(\aP^2)^2\big)\,\extd\varphi^2
}{eq:3d5}
means that in the second-order formulation, we have imposed the boundary conditions
\eq{
h_{\mu\nu}\,\extd x^\mu\extd x^\nu=\extd r^2 + r^2\,\extd\varphi^2 + \mathcal{O}(r)\,\extd\varphi^2 + \mathcal{O}(1)\,\extd r\extd\varphi\,.
}{eq:gettingolder}
The clock 1-form is given by
\eq{
\tau = \mu\,\extd t + \big(\aH -\aC^1 r\big)\,\extd\varphi
}{eq:3d6}
which through the defining relations $v^\mu \tau _\mu =-1$, $v^\mu e_\mu ^a=0$ determines the Carrollian vector field
\eq{
v^\mu\,\partial_\mu = -\frac{1}{\mu}\,\partial_t\,.
}{eq:3d7}
The postcarrollian 1-form 
\eq{
m^a_\mu\,\extd x^\mu=\big(\aS\,r\,\delta^a_1+\aM^a\big)\,\extd\varphi
}{eq:3d8}
has vanishing contraction with the Carrollian vector field \eqref{eq:3d7}. 

Parametrizing the boundary condition preserving gauge transformations \eqref{eq:hostage} by $\lambda =b^{-1}\varepsilon b$ one finds the conditions $\partial_t\varepsilon _I=0$ except for $\epsM$ which satisfies $\partial_t \epsM^a -\mu\,\epsilon^a{}_b\,\epsC^b=0$. Additionally, there is the constraint $\partial_\varphi \epsJ =0$ from the requirements in the boundary connection \eqref{eq:3d2} that the $J$-component is fixed to $1$.

The variation of the boundary charges 
\eq{
\delta Q[\varepsilon ] = \frac{k}{2\pi}\,\oint\extd\varphi\,\langle \varepsilon ,\,\delta a_\varphi\rangle
}{eq:halleyscomet}
can be functionally integrated for the boundary conditions \eqref{eq:3d2}, yielding 
\eq{
Q[\varepsilon ] = \frac{k}{2\pi}\,\oint\extd\varphi\,\big(\epsS\,\aH + \epsM{}_{b}\,\aC^b+\epsG{}_{b}\,\aP^b+\epsJ\,\aL+\epsP{}_{b}\,\aG^b+\epsC{}_{b}\,\aM^b+\epsH\,\aS\big)\,.
}{eq:notmyresponsibility}

The on-shell conditions $\partial_ta_\varphi-\partial _\varphi a_t+[a_t,\,a_\varphi]=0$ imply
\eq{
\partial_t \aH = \partial_\varphi\mu\qquad\qquad \partial_t \aM^a-\mu\,\epsilon^a{}_b\,\aC^b = 0\qquad \qquad \partial_t (\aP^a,\aC^a,\aL,\aG^a,\aS)=0
}{eq:oceaneyes}
which in turn establishes the conservation of the canonical boundary charges
\eq{
\partial_t Q[\varepsilon ] = \frac{k}{2\pi}\oint\extd\varphi\,\big(\partial_t \aH\epsS+\partial_t \aM^b\epsC{}_{b}+\partial_t\epsM{}_{b}\aC^b\big) = \frac{k}{2\pi}\oint\extd\varphi\,\partial_\varphi\mu\,\epsS
}{eq:bittersuite}
in the sense that the right-hand side is state-independent, i.e., $\partial_t\delta Q = 0$ as a consequence of the on-shell conditions \eqref{eq:oceaneyes}. For the asymptotic symmetry algebra, it is useful to absorb the time-dependence by redefining\footnote{%
Algebraically, we exploit the automorphism $C_n^a\to C_n^a-\nu\,\epsilon^a{}_b\,M_n^b$ of the asymptotic symmetry algebra \eqref{eq:oxytocin}.
}
\eq{
\epsM^a(t,\,\varphi)\to\epsM^a(\varphi)+\epsilon^a{}_b\,\epsC^b(\varphi)\,\nu(t,\,\varphi)\qquad\qquad \aM^a(t,\,\varphi)\to\aM^a(\varphi)+\epsilon^a{}_b\,\aC^b(\varphi)\,\nu(t,\,\varphi)
}{eq:wishyouweregay}
where $\nu(t,\,\varphi):=\int^t\!\mu(t^\prime,\varphi)\extd t^\prime$.

Introducing Fourier modes for the state-dependent functions and transformation parameters
\begin{subequations}
\label{eq:nda}
\begin{align}
P_n^a&=\frac{k}{2\pi}\oint\extd\varphi\,e^{in\varphi}\,\aG^a(\varphi) & C_n^a &= \frac{k}{2\pi}\oint\extd\varphi\,e^{in\varphi}\,\aM^a(\varphi)\\ 
J &= \frac{k}{2\pi}\oint\extd\varphi\,\aL(\varphi) & H_n &= \frac{k}{2\pi}\oint\extd\varphi\,e^{in\varphi}\aS(\varphi)\\
M_n^a&=\frac{k}{2\pi}\oint\extd\varphi\,e^{in\varphi}\,\aC^a(\varphi) & G_n^a &= \frac{k}{2\pi}\oint\extd\varphi\,e^{in\varphi}\,\aP^a(\varphi)\\
S_n &= \frac{k}{2\pi}\oint\extd\varphi\,e^{in\varphi}\,\aH(\varphi) & 
\varepsilon_{I\,n} &= \frac{1}{2\pi}\,\oint\extd\varphi\,e^{in\varphi}\,\varepsilon_I(\varphi)
\end{align}
\end{subequations}
yields the mode expansion of the postcarrollian asymptotic symmetry algebra via the relation $\delta_{\varepsilon_1}Q[\varepsilon_2]=\{Q[\varepsilon_1],\,Q[\varepsilon_2]\}$ (see, e.g., \cite{Grumiller:2022qhx})
\begin{subequations}
\label{eq:oxytocin}
\begin{align}
   \{J,\,P_n^a\} &= \epsilon_b{}^a\,P_n^b & \{J,\,C_n^a\} &= \epsilon_b{}^a\,C_n^b \\
   \{J,\,M_n^a\} &= \epsilon_b{}^a\,M_n^b & \{J,\,G_n^a\} &= \epsilon_b{}^a\,G_n^b  \\
   \{P_n^a,\,C_m^b\} &= -\epsilon^{ab}\,H_{n+m} & \{C_n^a,\,C_m^b\} &= \epsilon^{ab}\,S_{n+m} \\  
   \{P_n^a,\,G_m^b\} &= \big(in\,\delta^{ab}-\epsilon^{ab}\big)\,L\,\delta_{n+m,\,0}=\{M_n^a,\,C_m^b\}  & \{S_n,\,C_m^a\} &= \epsilon_b{}^a{}\,G^b_{n+m} \\  
    \{C_n^a,\,H_m\} &= \epsilon_b{}^a\,M^b_{n+m} =\{S_n,\,P_m^a\} & \{H_n,\,S_m\} &= in\,L\,\delta_{n+m,\,0}\,.
\end{align}
\end{subequations}

The postcarrollian asymptotic symmetry algebra \eqref{eq:oxytocin} features a central extension, which we labelled by the generator $L$ so that the zero-mode subalgebra (which coincides with the wedge subalgebra) matches with the postcarrollian algebra \eqref{PostCalgebra3D}. 

\subsection{Postcarrollian gravity in 4d}\label{se:6.2}

A 4d postcarrollian gravity action is defined as a MacDowell--Mansouri-like action,
\begin{equation}\label{4daction}
I_{4d}=-\frac{\kappa}{2\Lambda} \int \left\langle
F \wedge F 
\right\rangle
\end{equation}
invariant under the postcarrollian extension of the homogeneous Carroll symmetry, namely, the subalgebra of \eqref{PostCalgebra} spanned by the generators $\{B_a,J_{ab},K_a,S_{ab}\}$. This subalgebra admits the following non-degenerate invariant bilinear form ($\rho_1\neq 0$)
\begin{equation}\label{invtensor4d}
\left\langle B_a , J_{cd}
\right\rangle =\rho_0 \epsilon_{abc}\qquad \qquad 
\left\langle B_a , S_{cd}
\right\rangle = \rho_1 \epsilon_{abc}= \left\langle K_a , J_{cd}
\right\rangle ~.
\end{equation}

In 4d, the field strengh $F=\dd A+A\wedge A$ associated to the postcarrollian connection \eqref{connectionA},
\begin{align}
F&=T H+ T^a P_a
+\left(R^a(B)-\Lambda \tau\wedge e^a \right)B_a
+\frac12\left(R^{ab}(J)-\Lambda e^a\wedge e^b\right)J_{ab}
\\
&+\Theta L+ \Theta^a M_a
+\left(R^a(K)- \Lambda \tau\wedge m^a - \Lambda l\wedge e^a \right)K_a
+\frac12\left(R^{ab}(S)-2\Lambda e^{[a}\wedge m^{b]}\right)S_{ab} \nonumber 
\end{align}
contains the Carrollian torsion forms $T$, $T^a$ and the postcarrollian torsion forms $\Theta$, $\Theta^a$,
\begin{subequations}
\begin{align}
   T&=\dd \tau +\omega^a \wedge e_a  & 
    T^a&=De^a \label{TorsionT} \\ 
   \Theta &=\dd l+\omega^a \wedge m_a + \sigma^a \wedge e_a &
    \Theta^a &=D m^a
    +\sigma^a{}_b \wedge e^b + \omega^a \wedge \tau \label{TorsionTheta}
\end{align}
\end{subequations}
whereas the Carrollian curatures $R^a(B)$, $R^{ab}(J)$ and the postcarrollian curvature two-forms $R^a(K)$, $R^{ab}(S)$ are given by
\begin{subequations}\label{curvatures4d}
\begin{align}
    R^a(B)&=D \omega ^a &
    R^{ab}(J)&=\dd \omega ^{ab}+\omega ^a{}_c\wedge \omega ^{cb}\\
    R^a(K)&= D \sigma ^a+\sigma ^a{}_b\wedge \omega ^b &
    R^{ab}(S)&=D\sigma ^{ab}+\omega ^a\wedge \omega ^b
\end{align}
\end{subequations}
where we have introduced an $SO(3)$-covariant derivative $D$ acting on $SO(3)$-tensor-valued differential forms as
\begin{equation}\label{covDso3}
DX^{a_1\cdots a_s}=\dd X^{a_1\cdots a_s}+ \omega^{a_1}{}_b \wedge X^{b a_2\cdots a_s} + \omega^{a_2}{}_b \wedge X^{a_1 b a_3\cdots a_s} +\cdots +
\omega^{a_s}{}_b \wedge X^{a_1\cdots a_{s-1}b}~.
\end{equation}

The torsion and curvature forms satisfy Bianchi identities $\dd F+A\wedge F-F\wedge A=0$,
\begin{subequations}
    \label{eq:bianchi}
\begin{align}
    \dd T+\omega _a\wedge T^a&=e_a\wedge R^a (B)\\
    DT^a&=R^a{}_b(J)\wedge e^b\\
    \dd \Theta +\omega _a\wedge \Theta ^a+\sigma _a\wedge T^a&=m_a\wedge R^a(B)+e_a\wedge R^a (K)\\
    D\Theta ^a+\sigma ^a{}_b\wedge T^b+\omega_a\wedge T&=\tau \wedge R^a (B)+R^a{}_b(J)\wedge m^b+R^a{}_b(S)\wedge e^b\\
    DR^a(B) &=R^a{}_b(J)\wedge \omega ^b \\
    DR^{ab}(J)&=0\\
    D R^a(K)&=R^a{}_b(J)\wedge \sigma ^b+R^a{}_b(S)\wedge \omega ^b-\sigma^a{}_b\wedge R^b(B)\\
    DR^{ab}(S)&=2R^{[a}(B)\wedge \omega ^{b]}+2R_c{}^{[a}(J)\wedge \sigma ^{b]c} ~.
\end{align}
\end{subequations}

The gauge transformations for the Carrollian and postcarrollian gauge fields follow from the transformation rule of the connection one-form, $\delta A= \dd \lambda + [A,\lambda]$, where $\lambda$ is a local parameter taking values in the postcarrollian algebra \eqref{PostCalgebra}, i.e.,
\begin{align}
    \lambda =\xi H +\xi ^aP_a+\zeta L+\zeta ^aM_a+\theta ^aB_a+\frac{1}{2}\theta ^{ab}J_{ab}+\eta ^aK_a+\frac{1}{2}\eta ^{ab}S_{ab}~.
\end{align}
The gauge fields transform in the adjoint representation as
\begin{subequations}
\begin{align}
    \delta _\lambda \tau &=\dd\xi +\omega ^a\xi _a-\theta _ae^a \\
    \delta _\lambda e^a&=D\xi ^a-\theta ^a{}_be^b\\
    \delta _\lambda l&=\dd \zeta +\omega ^a\zeta _a-\theta _a m^a+\sigma ^a\xi _a-\eta _ae^a\\
    \delta _\lambda m^a&=D\zeta ^a-\theta ^a{}_bm^b+\sigma ^a{}_b\xi ^b-\eta ^a{}_be^b+\omega ^a\xi -\theta ^a\tau \\
    \delta _\lambda \omega ^a&=D\theta ^a-\theta ^a{}_b\omega ^b+\Lambda \big(e^a\xi -\xi ^a\tau \big)\\
    \delta _\lambda \omega ^{ab}&=D \theta ^{ab}-2\Lambda e^{[a}\xi ^{b]}\\
    \delta _\lambda \sigma ^a&=D\eta ^a-\theta ^a{}_b\sigma ^b+\sigma ^a{}_b\theta ^b-\eta ^a{}_b\omega ^b +\Lambda \big(m^a\xi -\zeta ^a\tau +e^a\zeta -\xi ^al\big) \\
    \delta _\lambda \sigma ^{ab}&=D \eta ^{ab}-2\theta ^{c[a}\sigma ^{b]}{}_c+2\omega ^{[a}\theta ^{b]}-\Lambda \big(e^{[a}\zeta ^{b]}-m^{[a}\xi ^{b]}\big)\,.
\end{align}
\end{subequations}

Using \eqref{invtensor4d} and \eqref{curvatures4d}, up to boundary terms the action \eqref{4daction} takes the form
\begin{equation}
I_{4d}= \rho_0\, I_{\textrm{\tiny (A)dSC}}+ \rho_1\, I_{\textrm{\tiny PC}}~.
\end{equation}
The term along $\rho_0$ is the  magnetic (A)dS Carroll gravity action
\begin{equation}
I_{\textrm{\tiny (A)dSC}}= \kappa \int \epsilon_{abc} \left( \tau\wedge  e^a \wedge R^{bc}(J) + e^a\wedge e^b \wedge R^c(B)- \Lambda \tau\wedge e^a \wedge e^b \wedge e^c\right)~.
\end{equation}
The $\Lambda=0$ limit of this action was introduced in \cite{Bergshoeff:2017btm} by gauging the Carroll symmetry, and was later shown in \cite{Campoleoni:2022ebj} to be equivalent to magnetic Carroll gravity \cite{Henneaux:2021yzg}.
The term along $\rho_1$ is the magnetic postcarrollian gravity action
\begin{align}\label{PCaction4D}
    I_{\textrm{\tiny PC}}&=\kappa \int
     \epsilon _{abc}\Big(\tau \wedge e^a\wedge R^{bc}(S)+\tau \wedge m^a\wedge R^{bc}(J)+l\wedge e^a\wedge R^{bc}(J)\\
    &+e^a\wedge e^b\wedge R^c(K)+2e^a\wedge m^b\wedge R^c(B)+\Lambda \left(3\tau \wedge m^a+l\wedge e^a\right)\wedge e^b\wedge e^c\Big) ~.
\end{align}
This action is obtained from the Einstein--Cartan action for gravity with cosmological constant as the sub-leading correction in an expansion in powers of $c^2$, where the Lorentzian frame-fields and connections expand as in \eqref{expansionRelFields}.

As in the 3d case, the field equations associated with the postcarrollian fields reproduce the field equations known from magnetic Carroll gravity. In particular, this leads to the vanishing of the Carrollian torsion forms \eqref{TorsionT}. These constraints allow for solving the spin connections in terms of the vielbein as
\begin{equation}\label{solomegas}
\omega^{ab}_\mu= 2 e^{[a|\nu} \partial_{[\mu} e^{b]}_{\nu]}
- e^{a\nu} e^{b\rho} e_{c\mu}\partial_{[\nu} e^c_{\rho]}
\qquad\quad
\omega^a_\mu= -e^{a\nu}\left( v^\rho \tau_\mu \partial_{[\nu}\tau_{\rho]}
+\partial_{[\mu}\tau_{\nu]}\right)+W^a{}_b e^b_\mu
\end{equation}
where $W^{ab}=\omega^{(a}_\mu e^{b)\mu}$ is a symmetric tensor encoding the part of the Carroll boost connection that is not determined by the Carrollian torsion constraints \cite{Bergshoeff:2017btm}. Here we used the definition of the inverse frame fields $v^\mu, e_a^\mu $
\begin{align}
    v^\mu \tau _\mu &=-1 & \delta ^\mu _\nu &=-v^\mu \tau _\nu +e^\mu _ae^a_\nu & v^\mu e_\mu ^a&=0 & e^\mu _a\tau _\mu &=0 ~.
\end{align}
Similarly, the variations of the postcarrollian action with respect to the spin connections yield the vanishing of the postcarrollian torsion two-forms \eqref{TorsionTheta}. The postcarrollian torsion constraints allow to solve the fields $\sigma^a$ and $\sigma^{ab}$, and fix the tensor $W^{ab}$ in terms of the postcarrollian field $m^a$
\begin{subequations}\label{solsigmas}
\begin{align}
\sigma^{ab}_\mu&= 2 e^{[a|\nu} D_{[\mu} m^{b]}_{\nu]}
- e^{a\nu} e^{b\rho} e_{c\mu}D_{[\nu} m^c_{\rho]}
- e^{[a|\nu}\omega^{b]}_\nu \tau_\mu
\\
\sigma^a_\mu&= e^{a\nu}\left(- v^\rho \tau_\mu +\delta^\rho_\mu\right)\left(\partial_{[\nu} l_{\rho]}+\omega_{a[\nu}m^a_{\rho]}\right)+V^a{}_b e^b_\mu
\\
W^{ab}&=2D_{[\mu} m^{(a}_{\nu]}e^{b)\mu} v^\nu~.
\end{align}
\end{subequations}
As in the previous case, $V^{ab}=\sigma^{(a}_\mu e^{b)\mu}$ is a symmetric tensor containing the part of $\sigma^{ab}$ that remains independent of the Carrollian frame-fields and their postcarrollian corrections. This situation resembles the Galilean case \cite{Andringa:2010it}, where the Bargmann extension allows to fix the components of the spin connection that were not determined by the Galilean torsion constraints. Indeed, in the context of non-relativistic expansions, the mass generator in the Bargmann algebra corresponds to a postgalilean correction. Replacing the solutions \eqref{solomegas} and \eqref{solsigmas} in the postcarrollian action \eqref{PCaction4D} leads to a second order formulation of the theory. Similarly to what happens in the case of magnetic Carroll gravity \cite{Bergshoeff:2017btm}, the contribution of the curvature two-form $R^a(K)$ yields term
\begin{equation}
\epsilon_{abc} e^a\wedge e^b\wedge D\sigma^c \,\sim\, V^{ab} \left(e_a^\mu e_b^\nu-\delta_{ab} e_c^\mu e^{c\nu} \right)K_{\mu\nu}\,\det (\tau _\alpha ,e_\alpha ^d) \dd ^4 x  
\end{equation}
where $K_{\mu\nu}=-\frac12 \mathcal L_v (e^c_\mu e_{c\nu})$ is the extrinsic curvature. Therefore $V_{ab}$ is a Lagrange multiplier that enforces the extrinsic curvature constraint $K_{\mu\nu}=0$.


\section{Discussion and Outlook}\label{se:7}

We finish with a few remarks and further points worth pursuing. First and foremost, postcarrollian corrections render the Hamiltonian $H$ no longer central so that the defining Carrollian property \eqref{eq:intro1} no longer holds. Physically, this means that boosting a postcarrollian state, in general, changes its energy so that postcarrollian descendants of a parent state no longer are ``soft''. This can provide a controlled cutoff on the spectrum of soft excitations (see the discussion in \cite{Bagchi:2022iqb} and Refs.~therein for why such a cutoff can be desirable).

As stated initially, the postcarrollian expansion we performed is around the magnetic theory and as such should contain corrections to Carroll black holes \cite{Ecker:2023uwm} away from the strict $c\to 0$ limit. Once additional corrections are taken into account, there should be a resummation that leads back to the full Lorentzian theory. It could be interesting to see how this works exactly. 

Remarkably, there are two distinct solution sectors for postcarrollian 2d dilaton gravity discussed in section \ref{se:4}, one of which leads to Carroll black hole solutions while the other yields cosmological solutions. Since this is qualitatively different from the leading order Carrollian theory, it could be rewarding to understand physically better why the postcarrollian corrections introduce this dichotomy. 
 
Studies of asymptotic symmetries of 2d, 3d, and 4d Carroll gravity show that non-trivial Drinfeld--Sokolov reductions exist \cite{Bergshoeff:2016soe,Grumiller:2020elf,Perez:2021abf,Perez:2022jpr,Aviles:2025ygw}, which provides encouragement to search for such symmetries in postcarrollian gravity. In section \ref{se:5} we found the twisted warped boundary action for the postcarrollian JT model and a more mundane boundary action for the postcarrollian CGHS model. It could be worthwhile to investigate alternatives to the boundary constraints we imposed. 

Another open task is to holographically renormalize the postcarrollian action \eqref{eq:action}. We suspect that this can be done by analogy to the Carrollian case \cite{Ecker:2024czh} and the Lorentzian case \cite{Grumiller:2007ju}.

In 3d and 4d, so far we only wrote down the models but did not investigate the solution space apart from the spherically symmetric sector, which is effectively again a postcarrollian 2d dilaton gravity theory. While finding solutions as general as in 2d could turn out to be as difficult as in fully fledged Einstein gravity, one could take a first step away from spherical symmetry and consider axisymmetric configurations. 

An explicit example for studying the postcarrollian corrections to a 3d Carroll geometry would be to take the recently found solution of the 3d AdS Carroll theory \cite{Aviles:2025ygw}. It already solves the leading order equations \eqref{eq:3d_LO_equations}, so it remains to figure out how the postcarrollian corrections can be solved for by \eqref{eq:eomPC}. It is interesting that the wedge subalgebra of the asymptotic symmetry algebra found in \cite{Aviles:2025ygw} is given by the subalgebra of postcarrollian symmetry \eqref{PostCalgebra3D} spanned by the generators $\{H,P_a,J,C_a,S,M_a\}$.

Finally, the boundary conditions imposed in section \ref{se:6.1.3} are just one example of Brown--Henneaux-inspired boundary conditions in postcarrollian gravity. They led to the asymptotic symmetry algebra \eqref{eq:oxytocin}, which features two independent central extensions and has the postcarrollian algebra \eqref{PostCalgebra3D} as wedge subalgebra. It would be gratifying to analyze more comprehensively the menagerie of possible boundary conditions, their associated asymptotic symmetries, and their central extensions, some of which may engender cardyological microstate counting of postcarrollian black hole states.


\section*{Acknowledgements}

We thank the participants of the Erwin-Schr\"odinger Institute (ESI) workshop ``Carroll physics \& holography'' for numerous discussions.
D.G.~is grateful to the organizers and participants of the Workshop on Quantum Gravity, Strings and the Swampland in Corfu, September 2024, in particular, to Jan Rosseel and George Zoupanos. The talk delivered at this Workshop was on Carroll swiftons \cite{Ecker:2024czx} and not on the content of this proceedings contribution, which contains new results.
F.E.~and D.G.~were supported by the Austrian Science Fund (FWF) [Grants DOI:
\href{https://www.fwf.ac.at/forschungsradar/10.55776/P32581}{10.55776/P32581}, DOI: \href{https://www.fwf.ac.at/forschungsradar/10.55776/P33789}{10.55776/P33789}, and DOI: \href{https://www.fwf.ac.at/forschungsradar/10.55776/P36619}{10.55776/P36619}].
P.S.-R.~has been supported by a Young Scientist Training Program (YST) fellowship at the Asia Pacific Center for Theoretical Physics (APCTP) through the Science and Technology Promotion Fund and the Lottery Fund of the Korean Government. P.S.-R. has also been supported by the Korean local governments in Gyeongsangbuk-do Province and Pohang City. Part of this work emerged during the ESI workshop \href{https://www.esi.ac.at/events/e518/}{``Carroll physics \& holography''} in April 2024. All authors have benefited from the OeAD travel grant IN 04/2024, which allowed them to visit BITS Pilani in Goa in February 2024.


\begin{appendix}

\section{Baker--Campbell--Hausdorff identities}\label{app:A}

For the calculations in this work, we found three Baker--Campbell--Hausdorff identities useful. The first one,
\eq{
[X,\,Y] = \sigma Y \qquad\Rightarrow\qquad e^X\, e^Y = e^{e^\sigma\,Y}\, e^X
}{eq:app1}
is a well-known braiding identity and implies, for instance,
\eq{
e^{p\,P}\,e^{y\,L_+} = e^{e^{-p}\,y\,L_+}\,e^{p\,P}
}{eq:app2}
for the commutator $[L_+,\,P]=L_+$. 

The second one,
\eq{
[X,\,Y] = Z\qquad[Z,\,\cdot]=0\qquad\Rightarrow\qquad e^X\, e^Y = e^Y\, e^X\, e^Z
}{eq:app3}
is well known from the Heisenberg algebra and implies, for instance,
\eq{
e^{y\,L_+}\, e^{y_-\,L_-} = e^{y_-\,L_-}\, e^{y\,L_+}\, e^{2yy_-\,M}
}{eq:app4}
for the commutator $[L_+,\,L_-]=2M$.

The third one applies to the postcarrollian CGHS commutation relations
\eq{
[B,\,P]=H\qquad\qquad[B,\,H]=M\qquad\qquad[H,\,P]=Z\qquad\qquad[M,\,\cdot]=[Z,\,\cdot]=0
}{eq:app5}
and yields the identity
\eq{
e^{-b\,B}\,P\,e^{b\,B} = P-b\,H+\frac{b^2}{2}\,M\,.
}{eq:app6}
Moreover, applying \eqref{eq:app3} yields
\eq{
e^{-p\,P}\,H\,e^{p\,P}=H+p\,Z\qquad\qquad e^{-b\,B}\,H\,e^{b\,B}=H-b\,M\,.
}{eq:app7}

\end{appendix}


\providecommand{\href}[2]{#2}\begingroup\raggedright\endgroup



\begin{thebibliography}{10}

\addcontentsline{toc}{section}{References}

\bibitem{LevyLeblond1965}
J.-M. L\'evy-Leblond, ``{Une nouvelle limite non-relativiste du groupe de
  Poincar\'e},'' {\em A. Inst. Henri Poincar\'e III 1} (1965).

\bibitem{Gupta1966}
N.~D. SenGupta, ``On an analogue of the galilei group,'' {\em Il Nuovo Cimento
  A Series 10} {\bf 44} (1966), no.~2, 512--517.

\bibitem{Souriau:1973}
J.-M. Souriau, ``Ondes et radiations gravitationnelles,'' {\em Colloques
  Internationaux du CNRS} {\bf 220} (1973) 243--256.

\bibitem{Elbistan:2023qbp}
M.~Elbistan, P.~M. Zhang, and P.~A. Horvathy, ``{Memory effect \& Carroll
  symmetry, 50 years later},'' {\em Annals Phys.} {\bf 459} (2023) 169535,
  \href{http://www.arXiv.org/abs/2306.14271}{{\tt 2306.14271}}.

\bibitem{Henneaux:1979vn}
M.~Henneaux, ``{Geometry of Zero Signature Space-times},'' {\em Bull. Soc.
  Math. Belg.} {\bf 31} (1979) 47--63.

\bibitem{Henneaux:1982qpq}
M.~Henneaux, ``{Quantification hamiltonienne du champ de gravitation : une
  nouvelle approche},'' {\em Bulletin de la Classe des sciences} {\bf 68}
  (1982), no.~1, 940--971.

\bibitem{Damour:2002et}
T.~Damour, M.~Henneaux, and H.~Nicolai, ``{Cosmological billiards},'' {\em
  Class. Quant. Grav.} {\bf 20} (2003) R145--R200,
  \href{http://www.arXiv.org/abs/hep-th/0212256}{{\tt hep-th/0212256}}.

\bibitem{deBoer:2021jej}
J.~de~Boer, J.~Hartong, N.~A. Obers, W.~Sybesma, and S.~Vandoren, ``{Carroll
  Symmetry, Dark Energy and Inflation},'' {\em Front. in Phys.} {\bf 10} (2022)
  810405, \href{http://www.arXiv.org/abs/2110.02319}{{\tt 2110.02319}}.

\bibitem{Gibbons:2002tv}
G.~Gibbons, K.~Hashimoto, and P.~Yi, ``{Tachyon condensates, Carrollian
  contraction of Lorentz group, and fundamental strings},'' {\em JHEP} {\bf 09}
  (2002) 061, \href{http://www.arXiv.org/abs/hep-th/0209034}{{\tt
  hep-th/0209034}}.

\bibitem{Barnich:2006av}
G.~Barnich and G.~Comp{\`e}re, ``{Classical central extension for asymptotic
  symmetries at null infinity in three spacetime dimensions},'' {\em
  Class.Quant.Grav.} {\bf 24} (2007) F15--F23,
\href{http://www.arXiv.org/abs/gr-qc/0610130}{{\tt gr-qc/0610130}}.

\bibitem{Bagchi:2010zz}
A.~Bagchi, ``{Correspondence between Asymptotically Flat Spacetimes and
  Nonrelativistic Conformal Field Theories},'' {\em Phys.Rev.Lett.} {\bf 105}
  (2010)
171601.

\bibitem{Bagchi:2012yk}
A.~Bagchi, S.~Detournay, and D.~Grumiller, ``{Flat-Space Chiral Gravity},''
  {\em Phys.Rev.Lett.} {\bf 109} (2012) 151301,
\href{http://www.arXiv.org/abs/1208.1658}{{\tt 1208.1658}}.

\bibitem{Bagchi:2012xr}
A.~Bagchi, S.~Detournay, R.~Fareghbal, and J.~Simon, ``{Holography of 3d Flat
  Cosmological Horizons},'' {\em Phys. Rev. Lett.} {\bf 110} (2013) 141302,
\href{http://www.arXiv.org/abs/1208.4372}{{\tt 1208.4372}}.

\bibitem{Barnich:2012xq}
G.~Barnich, ``{Entropy of three-dimensional asymptotically flat cosmological
  solutions},'' {\em JHEP} {\bf 1210} (2012) 095,
\href{http://www.arXiv.org/abs/1208.4371}{{\tt 1208.4371}}.

\bibitem{Barnich:2013yka}
G.~Barnich and H.~A. Gonzalez, ``{Dual dynamics of three dimensional
  asymptotically flat Einstein gravity at null infinity},'' {\em JHEP} {\bf
  1305} (2013) 016,
\href{http://www.arXiv.org/abs/1303.1075}{{\tt 1303.1075}}.

\bibitem{Duval:2014uva}
C.~Duval, G.~W. Gibbons, and P.~A. Horvathy, ``{Conformal Carroll groups and
  BMS symmetry},'' {\em Class. Quant. Grav.} {\bf 31} (2014) 092001,
  \href{http://www.arXiv.org/abs/1402.5894}{{\tt 1402.5894}}.

\bibitem{Bagchi:2014iea}
A.~Bagchi, R.~Basu, D.~Grumiller, and M.~Riegler, ``{Entanglement entropy in
  Galilean conformal field theories and flat holography},'' {\em
  Phys.Rev.Lett.} {\bf 114} (2015), no.~11, 111602,
\href{http://www.arXiv.org/abs/1410.4089}{{\tt 1410.4089}}.

\bibitem{Donnay:2022aba}
L.~Donnay, A.~Fiorucci, Y.~Herfray, and R.~Ruzziconi, ``{Carrollian Perspective
  on Celestial Holography},'' {\em Phys. Rev. Lett.} {\bf 129} (2022), no.~7,
  071602, \href{http://www.arXiv.org/abs/2202.04702}{{\tt 2202.04702}}.

\bibitem{Bagchi:2022emh}
A.~Bagchi, S.~Banerjee, R.~Basu, and S.~Dutta, ``{Scattering Amplitudes:
  Celestial and Carrollian},'' {\em Phys. Rev. Lett.} {\bf 128} (2022), no.~24,
  241601, \href{http://www.arXiv.org/abs/2202.08438}{{\tt 2202.08438}}.

\bibitem{deBoer:2017ing}
J.~de~Boer, J.~Hartong, N.~A. Obers, W.~Sybesma, and S.~Vandoren, ``{Perfect
  Fluids},'' {\em SciPost Phys.} {\bf 5} (2018), no.~1, 003,
  \href{http://www.arXiv.org/abs/1710.04708}{{\tt 1710.04708}}.

\bibitem{Ciambelli:2018xat}
L.~Ciambelli, C.~Marteau, A.~C. Petkou, P.~M. Petropoulos, and K.~Siampos,
  ``{Covariant Galilean versus Carrollian hydrodynamics from relativistic
  fluids},'' {\em Class. Quant. Grav.} {\bf 35} (2018), no.~16, 165001,
  \href{http://www.arXiv.org/abs/1802.05286}{{\tt 1802.05286}}.

\bibitem{Penna:2018gfx}
R.~F. Penna, ``{Near-horizon Carroll symmetry and black hole Love numbers},''
  \href{http://www.arXiv.org/abs/1812.05643}{{\tt 1812.05643}}.

\bibitem{Donnay:2019jiz}
L.~Donnay and C.~Marteau, ``{Carrollian Physics at the Black Hole Horizon},''
  {\em Class. Quant. Grav.} {\bf 36} (2019), no.~16, 165002,
  \href{http://www.arXiv.org/abs/1903.09654}{{\tt 1903.09654}}.

\bibitem{Ciambelli:2019lap}
L.~Ciambelli, R.~G. Leigh, C.~Marteau, and P.~M. Petropoulos, ``{Carroll
  Structures, Null Geometry and Conformal Isometries},'' {\em Phys. Rev. D}
  {\bf 100} (2019), no.~4, 046010,
  \href{http://www.arXiv.org/abs/1905.02221}{{\tt 1905.02221}}.

\bibitem{Redondo-Yuste:2022czg}
J.~Redondo-Yuste and L.~Lehner, ``{Non-linear black hole dynamics and
  Carrollian fluids},'' {\em JHEP} {\bf 02} (2023) 240,
  \href{http://www.arXiv.org/abs/2212.06175}{{\tt 2212.06175}}.

\bibitem{Freidel:2022vjq}
L.~Freidel and P.~Jai-akson, ``{Carrollian hydrodynamics and symplectic
  structure on stretched horizons},'' {\em JHEP} {\bf 05} (2024) 135,
  \href{http://www.arXiv.org/abs/2211.06415}{{\tt 2211.06415}}.

\bibitem{Gray:2022svz}
F.~Gray, D.~Kubiznak, T.~R. Perche, and J.~Redondo-Yuste, ``{Carrollian motion
  in magnetized black hole horizons},'' {\em Phys. Rev. D} {\bf 107} (2023),
  no.~6, 064009, \href{http://www.arXiv.org/abs/2211.13695}{{\tt 2211.13695}}.

\bibitem{Ciambelli:2018wre}
L.~Ciambelli, C.~Marteau, A.~C. Petkou, P.~M. Petropoulos, and K.~Siampos,
  ``{Flat holography and Carrollian fluids},'' {\em JHEP} {\bf 07} (2018) 165,
  \href{http://www.arXiv.org/abs/1802.06809}{{\tt 1802.06809}}.

\bibitem{Figueroa-OFarrill:2021sxz}
J.~Figueroa-O'Farrill, E.~Have, S.~Prohazka, and J.~Salzer, ``{Carrollian and
  celestial spaces at infinity},'' {\em JHEP} {\bf 09} (2022) 007,
  \href{http://www.arXiv.org/abs/2112.03319}{{\tt 2112.03319}}.

\bibitem{Herfray:2021qmp}
Y.~Herfray, ``{Carrollian manifolds and null infinity: a view from Cartan
  geometry},'' {\em Class. Quant. Grav.} {\bf 39} (2022), no.~21, 215005,
  \href{http://www.arXiv.org/abs/2112.09048}{{\tt 2112.09048}}.

\bibitem{Mittal:2022ywl}
N.~Mittal, P.~M. Petropoulos, D.~Rivera-Betancour, and M.~Vilatte, ``{Ehlers,
  Carroll, charges and dual charges},'' {\em JHEP} {\bf 07} (2023) 065,
  \href{http://www.arXiv.org/abs/2212.14062}{{\tt 2212.14062}}.

\bibitem{Bidussi:2021nmp}
L.~Bidussi, J.~Hartong, E.~Have, J.~Musaeus, and S.~Prohazka, ``{Fractons,
  dipole symmetries and curved spacetime},'' {\em SciPost Phys.} {\bf 12}
  (2022), no.~6, 205, \href{http://www.arXiv.org/abs/2111.03668}{{\tt
  2111.03668}}.

\bibitem{Marsot:2022imf}
L.~Marsot, P.~M. Zhang, M.~Chernodub, and P.~A. Horvathy, ``{Hall effects in
  Carroll dynamics},'' {\em Phys. Rept.} {\bf 1028} (2023) 1--60,
  \href{http://www.arXiv.org/abs/2212.02360}{{\tt 2212.02360}}.

\bibitem{Figueroa-OFarrill:2023vbj}
J.~Figueroa-O'Farrill, A.~P\'erez, and S.~Prohazka, ``{Carroll/fracton
  particles and their correspondence},'' {\em JHEP} {\bf 06} (2023) 207,
  \href{http://www.arXiv.org/abs/2305.06730}{{\tt 2305.06730}}.

\bibitem{Rodriguez:2021tcz}
P.~Rodr\'\i{}guez, D.~Tempo, and R.~Troncoso, ``{Mapping relativistic to
  ultra/non-relativistic conformal symmetries in 2D and finite $
  \sqrt{T\overline{T}} $ deformations},'' {\em JHEP} {\bf 11} (2021) 133,
  \href{http://www.arXiv.org/abs/2106.09750}{{\tt 2106.09750}}.

\bibitem{Bagchi:2022eui}
A.~Bagchi, A.~Banerjee, R.~Basu, M.~Islam, and S.~Mondal, ``{Magic fermions:
  Carroll and flat bands},'' {\em JHEP} {\bf 03} (2023) 227,
  \href{http://www.arXiv.org/abs/2211.11640}{{\tt 2211.11640}}.

\bibitem{Bagchi:2015nca}
A.~Bagchi, S.~Chakrabortty, and P.~Parekh, ``{Tensionless Strings from
  Worldsheet Symmetries},'' {\em JHEP} {\bf 01} (2016) 158,
  \href{http://www.arXiv.org/abs/1507.04361}{{\tt 1507.04361}}.

\bibitem{Bagchi:2022iqb}
A.~Bagchi, D.~Grumiller, S.~Sheikh-Jabbari, and M.~M. Sheikh-Jabbari,
  ``{Horizon strings as 3D black hole microstates},'' {\em SciPost Phys.} {\bf
  15} (2023), no.~5, 210, \href{http://www.arXiv.org/abs/2210.10794}{{\tt
  2210.10794}}.

\bibitem{Bagchi:2023cfp}
A.~Bagchi, A.~Banerjee, J.~Hartong, E.~Have, K.~S. Kolekar, and M.~Mandlik,
  ``{Strings near black holes are Carrollian},'' {\em Phys. Rev. D} {\bf 110}
  (2024), no.~8, 086009, \href{http://www.arXiv.org/abs/2312.14240}{{\tt
  2312.14240}}.

\bibitem{Ecker:2023uwm}
F.~Ecker, D.~Grumiller, J.~Hartong, A.~P\'erez, S.~Prohazka, and R.~Troncoso,
  ``{Carroll black holes},'' {\em SciPost Phys.} {\bf 15} (2023), no.~6, 245,
  \href{http://www.arXiv.org/abs/2308.10947}{{\tt 2308.10947}}.

\bibitem{Baig:2023yaz}
S.~A. Baig, J.~Distler, A.~Karch, A.~Raz, and H.-Y. Sun, ``{Spacetime Subsystem
  Symmetries},'' \href{http://www.arXiv.org/abs/2303.15590}{{\tt 2303.15590}}.

\bibitem{Kasikci:2023tvs}
O.~Kasikci, M.~Ozkan, and Y.~Pang, ``{Carrollian origin of spacetime subsystem
  symmetry},'' {\em Phys. Rev. D} {\bf 108} (2023), no.~4, 045020,
  \href{http://www.arXiv.org/abs/2304.11331}{{\tt 2304.11331}}.

\bibitem{Ecker:2024czx}
F.~Ecker, D.~Grumiller, M.~Henneaux, and P.~Salgado-Rebolledo, ``{Carroll
  swiftons},'' {\em Phys. Rev. D} {\bf 110} (2024), no.~4, L041901,
  \href{http://www.arXiv.org/abs/2403.00544}{{\tt 2403.00544}}.

\bibitem{Bagchi:2024qsb}
A.~Bagchi, P.~Chakraborty, S.~Chakrabortty, S.~Fredenhagen, D.~Grumiller, and
  P.~Pandit, ``{Boundary Carrollian Conformal Field Theories and Open Null
  Strings},'' {\em Phys. Rev. Lett.} {\bf 134} (2025), no.~7, 071604,
  \href{http://www.arXiv.org/abs/2409.01094}{{\tt 2409.01094}}.

\bibitem{Bagchi:2024ikw}
A.~Bagchi, A.~Banerjee, S.~Mondal, and S.~Sarkar, ``{Carroll in Shallow
  Water},'' \href{http://www.arXiv.org/abs/2411.04190}{{\tt 2411.04190}}.

\bibitem{Biswas:2025dte}
S.~Biswas, A.~Dubey, S.~Mondal, A.~Banerjee, A.~Kundu, and A.~Bagchi,
  ``{Carroll at Phase Separation},''
  \href{http://www.arXiv.org/abs/2501.16426}{{\tt 2501.16426}}.

\bibitem{deBoer:2023fnj}
J.~de~Boer, J.~Hartong, N.~A. Obers, W.~Sybesma, and S.~Vandoren, ``{Carroll
  stories},'' {\em JHEP} {\bf 09} (2023) 148,
  \href{http://www.arXiv.org/abs/2307.06827}{{\tt 2307.06827}}.

\bibitem{Ciambelli:2023xqk}
L.~Ciambelli, ``{Dynamics of Carrollian scalar fields},'' {\em Class. Quant.
  Grav.} {\bf 41} (2024), no.~16, 165011,
  \href{http://www.arXiv.org/abs/2311.04113}{{\tt 2311.04113}}.

\bibitem{Bergshoeff:2024ilz}
E.~A. Bergshoeff, P.~Concha, O.~Fierro, E.~Rodr\'\i{}guez, and J.~Rosseel, ``{A
  Conformal Approach to Carroll Gravity},''
  \href{http://www.arXiv.org/abs/2412.17752}{{\tt 2412.17752}}.

\bibitem{Hartong:2015xda}
J.~Hartong, ``{Gauging the Carroll Algebra and Ultra-Relativistic Gravity},''
  {\em JHEP} {\bf 08} (2015) 069,
  \href{http://www.arXiv.org/abs/1505.05011}{{\tt 1505.05011}}.

\bibitem{Grumiller:2020elf}
D.~Grumiller, J.~Hartong, S.~Prohazka, and J.~Salzer, ``{Limits of JT
  gravity},'' {\em JHEP} {\bf 02} (2021) 134,
  \href{http://www.arXiv.org/abs/2011.13870}{{\tt 2011.13870}}.

\bibitem{Gomis:2020wxp}
J.~Gomis, D.~Hidalgo, and P.~Salgado-Rebolledo, ``{Non-relativistic and
  Carrollian limits of Jackiw-Teitelboim gravity},'' {\em JHEP} {\bf 05} (2021)
  162, \href{http://www.arXiv.org/abs/2011.15053}{{\tt 2011.15053}}.

\bibitem{Aviles:2022xyx}
L.~Avil\'es, J.~Gomis, D.~Hidalgo, and J.~Zanelli, ``{Electric/magnetic
  Newton-Hooke and Carroll Jackiw-Teitelboim gravity},'' {\em JHEP} {\bf 02}
  (2023) 061, \href{http://www.arXiv.org/abs/2211.03633}{{\tt 2211.03633}}.

\bibitem{Hansen:2021fxi}
D.~Hansen, N.~A. Obers, G.~Oling, and B.~T. S\o{}gaard, ``{Carroll Expansion of
  General Relativity},'' {\em SciPost Phys.} {\bf 13} (2022), no.~3, 055,
  \href{http://www.arXiv.org/abs/2112.12684}{{\tt 2112.12684}}.

\bibitem{Boulanger:2002bt}
N.~Boulanger, M.~Henneaux, and P.~van Nieuwenhuizen, ``{Conformal
  (super)gravities with several gravitons},'' {\em JHEP} {\bf 01} (2002) 035,
  \href{http://www.arXiv.org/abs/hep-th/0201023}{{\tt hep-th/0201023}}.

\bibitem{deAzcarraga:2002xi}
J.~A. de~Azcarraga, J.~M. Izquierdo, M.~Picon, and O.~Varela, ``{Generating Lie
  and gauge free differential (super)algebras by expanding Maurer-Cartan forms
  and Chern-Simons supergravity},'' {\em Nucl. Phys. B} {\bf 662} (2003)
  185--219, \href{http://www.arXiv.org/abs/hep-th/0212347}{{\tt
  hep-th/0212347}}.

\bibitem{Izaurieta:2006zz}
F.~Izaurieta, E.~Rodriguez, and P.~Salgado, ``{Expanding Lie (super)algebras
  through Abelian semigroups},'' {\em J. Math. Phys.} {\bf 47} (2006) 123512,
  \href{http://www.arXiv.org/abs/hep-th/0606215}{{\tt hep-th/0606215}}.

\bibitem{Bergshoeff:2019ctr}
E.~Bergshoeff, J.~M. Izquierdo, T.~Ort\'\i{}n, and L.~Romano, ``{Lie Algebra
  Expansions and Actions for Non-Relativistic Gravity},'' {\em JHEP} {\bf 08}
  (2019) 048, \href{http://www.arXiv.org/abs/1904.08304}{{\tt 1904.08304}}.

\bibitem{Gomis:2019nih}
J.~Gomis, A.~Kleinschmidt, J.~Palmkvist, and P.~Salgado-Rebolledo,
  ``{Newton-Hooke/Carrollian expansions of (A)dS and Chern-Simons gravity},''
  {\em JHEP} {\bf 02} (2020) 009,
  \href{http://www.arXiv.org/abs/1912.07564}{{\tt 1912.07564}}.

\bibitem{Bacry:1968zf}
H.~Bacry and J.~Levy-Leblond, ``{Possible kinematics},'' {\em J. Math. Phys.}
  {\bf 9} (1968) 1605--1614.

\bibitem{Matulich:2019cdo}
J.~Matulich, S.~Prohazka, and J.~Salzer, ``{Limits of three-dimensional gravity
  and metric kinematical Lie algebras in any dimension},'' {\em JHEP} {\bf 07}
  (2019) 118, \href{http://www.arXiv.org/abs/1903.09165}{{\tt 1903.09165}}.

\bibitem{Cangemi:1992bj}
D.~Cangemi and R.~Jackiw, ``Gauge invariant formulations of lineal gravity,''
  {\em Phys. Rev. Lett.} {\bf 69} (1992) 233--236,
\href{http://arXiv.org/abs/hep-th/9203056}{{\tt hep-th/9203056}}.

\bibitem{Afshar:2019axx}
H.~Afshar, H.~A. Gonz\'alez, D.~Grumiller, and D.~Vassilevich, ``{Flat space
  holography and the complex Sachdev-Ye-Kitaev model},'' {\em Phys. Rev. D}
  {\bf 101} (2020), no.~8, 086024,
  \href{http://www.arXiv.org/abs/1911.05739}{{\tt 1911.05739}}.

\bibitem{Bergshoeff:2017btm}
E.~Bergshoeff, J.~Gomis, B.~Rollier, J.~Rosseel, and T.~ter Veldhuis,
  ``{Carroll versus Galilei Gravity},'' {\em JHEP} {\bf 03} (2017) 165,
  \href{http://www.arXiv.org/abs/1701.06156}{{\tt 1701.06156}}.

\bibitem{Campoleoni:2022ebj}
A.~Campoleoni, M.~Henneaux, S.~Pekar, A.~P\'erez, and P.~Salgado-Rebolledo,
  ``{Magnetic Carrollian gravity from the Carroll algebra},'' {\em JHEP} {\bf
  09} (2022) 127, \href{http://www.arXiv.org/abs/2207.14167}{{\tt 2207.14167}}.

\bibitem{Callan:1992rs}
C.~G. Callan, Jr., S.~B. Giddings, J.~A. Harvey, and A.~Strominger,
  ``Evanescent black holes,'' {\em Phys. Rev.} {\bf D45} (1992) 1005--1009,
\href{http://www.arXiv.org/abs/hep-th/9111056}{{\tt hep-th/9111056}}.

\bibitem{Grumiller:2021cwg}
D.~Grumiller, R.~Ruzziconi, and C.~Zwikel, ``{Generalized dilaton gravity in
  2d},'' {\em SciPost Phys.} {\bf 12} (2022), no.~1, 032,
  \href{http://www.arXiv.org/abs/2109.03266}{{\tt 2109.03266}}.

\bibitem{Ikeda:1993fh}
N.~Ikeda, ``{Two-dimensional gravity and nonlinear gauge theory},'' {\em Annals
  Phys.} {\bf 235} (1994) 435--464,
\href{http://www.arXiv.org/abs/hep-th/9312059}{{\tt hep-th/9312059}}.

\bibitem{Schaller:1994es}
P.~Schaller and T.~Strobl, ``Poisson structure induced (topological) field
  theories,'' {\em Mod. Phys. Lett.} {\bf A9} (1994) 3129--3136,
\href{http://arXiv.org/abs/hep-th/9405110}{{\tt hep-th/9405110}}.

\bibitem{Grumiller:2002nm}
D.~Grumiller, W.~Kummer, and D.~V. Vassilevich, ``Dilaton gravity in two
  dimensions,'' {\em Phys. Rept.} {\bf 369} (2002) 327--429,
\href{http://arXiv.org/abs/hep-th/0204253}{{\tt hep-th/0204253}}.

\bibitem{Bergshoeff:2016soe}
E.~Bergshoeff, D.~Grumiller, S.~Prohazka, and J.~Rosseel, ``{Three-dimensional
  Spin-3 Theories Based on General Kinematical Algebras},'' {\em JHEP} {\bf 01}
  (2017) 114,
\href{http://www.arXiv.org/abs/1612.02277}{{\tt 1612.02277}}.

\bibitem{Gonzalez:2018enk}
H.~A. Gonz\'alez, D.~Grumiller, and J.~Salzer, ``{Towards a bulk description of
  higher spin SYK},'' {\em JHEP} {\bf 05} (2018) 083,
  \href{http://www.arXiv.org/abs/1802.01562}{{\tt 1802.01562}}.

\bibitem{Grumiller:2017qao}
D.~Grumiller, R.~McNees, J.~Salzer, C.~Valc\'arcel, and D.~Vassilevich,
  ``{Menagerie of AdS$_{2}$ boundary conditions},'' {\em JHEP} {\bf 10} (2017)
  203, \href{http://www.arXiv.org/abs/1708.08471}{{\tt 1708.08471}}.

\bibitem{Mertens:2017mtv}
T.~G. Mertens, G.~J. Turiaci, and H.~L. Verlinde, ``{Solving the Schwarzian via
  the Conformal Bootstrap},'' {\em JHEP} {\bf 08} (2017) 136,
  \href{http://www.arXiv.org/abs/1705.08408}{{\tt 1705.08408}}.

\bibitem{Mertens:2018fds}
T.~G. Mertens, ``{The Schwarzian theory — origins},'' {\em JHEP} {\bf 05}
  (2018) 036,
\href{http://www.arXiv.org/abs/1801.09605}{{\tt 1801.09605}}.

\bibitem{Witten:1989hf}
E.~Witten, ``Quantum field theory and the {J}ones polynomial,'' {\em Commun.
  Math. Phys.} {\bf 121} (1989)
351.

\bibitem{Elitzur:1989nr}
S.~Elitzur, G.~W. Moore, A.~Schwimmer, and N.~Seiberg, ``{Remarks on the
  Canonical Quantization of the Chern-Simons-Witten Theory},'' {\em Nucl.
  Phys.} {\bf B326} (1989)
108--134.

\bibitem{Brown:1986nw}
J.~D. Brown and M.~Henneaux, ``{Central Charges in the Canonical Realization of
  Asymptotic Symmetries: An Example from Three-Dimensional Gravity},'' {\em
  Commun. Math. Phys.} {\bf 104} (1986)
207--226.

\bibitem{Coussaert:1995zp}
O.~Coussaert, M.~Henneaux, and P.~van Driel, ``{The Asymptotic dynamics of
  three-dimensional Einstein gravity with a negative cosmological constant},''
  {\em Class.Quant.Grav.} {\bf 12} (1995) 2961--2966,
\href{http://www.arXiv.org/abs/gr-qc/9506019}{{\tt gr-qc/9506019}}.

\bibitem{Maldacena:2016hyu}
J.~Maldacena and D.~Stanford, ``{Remarks on the Sachdev-Ye-Kitaev model},''
  {\em Phys. Rev. D} {\bf 94} (2016), no.~10, 106002,
  \href{http://www.arXiv.org/abs/1604.07818}{{\tt 1604.07818}}.

\bibitem{Sarosi:2017ykf}
G.~S\'arosi, ``{AdS$_{2}$ holography and the SYK model},'' {\em PoS} {\bf
  Modave2017} (2018) 001,
\href{http://www.arXiv.org/abs/1711.08482}{{\tt 1711.08482}}.

\bibitem{Gu:2019jub}
Y.~Gu, A.~Kitaev, S.~Sachdev, and G.~Tarnopolsky, ``{Notes on the complex
  Sachdev-Ye-Kitaev model},'' {\em JHEP} {\bf 02} (2020) 157,
  \href{http://www.arXiv.org/abs/1910.14099}{{\tt 1910.14099}}.

\bibitem{Galajinsky:2019lak}
A.~Galajinsky, ``{Schwarzian mechanics via nonlinear realizations},'' {\em
  Phys. Lett. B} {\bf 795} (2019) 277--280,
  \href{http://www.arXiv.org/abs/1905.01935}{{\tt 1905.01935}}.

\bibitem{Afshar:2015wjm}
H.~Afshar, S.~Detournay, D.~Grumiller, and B.~Oblak, ``{Near-Horizon Geometry
  and Warped Conformal Symmetry},'' {\em JHEP} {\bf 03} (2016) 187,
\href{http://www.arXiv.org/abs/1512.08233}{{\tt 1512.08233}}.

\bibitem{Afshar:2019tvp}
H.~R. Afshar, ``{Warped Schwarzian theory},'' {\em JHEP} {\bf 02} (2020) 126,
  \href{http://www.arXiv.org/abs/1908.08089}{{\tt 1908.08089}}.

\bibitem{Afshar:2021qvi}
H.~Afshar and B.~Oblak, ``{Flat JT gravity and the BMS-Schwarzian},'' {\em
  JHEP} {\bf 11} (2022) 172, \href{http://www.arXiv.org/abs/2112.14609}{{\tt
  2112.14609}}.

\bibitem{Ravera:2019ize}
L.~Ravera, ``{AdS Carroll Chern-Simons supergravity in 2 + 1 dimensions and its
  flat limit},'' {\em Phys. Lett. B} {\bf 795} (2019) 331--338,
  \href{http://www.arXiv.org/abs/1905.00766}{{\tt 1905.00766}}.

 \bibitem{Concha:2024tcu}
P.~Concha, E.~Rodr\'\i{}guez, and S.~Salgado, ``{3D Carrollian gravity from 2D
  Euclidean symmetry},'' \href{http://www.arXiv.org/abs/2501.00205}{{\tt
  2501.00205}}.


\bibitem{Hansen:2020pqs}
D.~Hansen, J.~Hartong, and N.~A. Obers, ``{Non-Relativistic Gravity and its
  Coupling to Matter},'' {\em JHEP} {\bf 06} (2020) 145,
  \href{http://www.arXiv.org/abs/2001.10277}{{\tt 2001.10277}}.

\bibitem{Achucarro:1993fd}
A.~Achucarro and M.~E. Ortiz, ``Relating black holes in two-dimensions and
  three- dimensions,'' {\em Phys. Rev.} {\bf D48} (1993) 3600--3605,
\href{http://www.arXiv.org/abs/hep-th/9304068}{{\tt hep-th/9304068}}.

\bibitem{Grumiller:2016pqb}
D.~Grumiller and M.~Riegler, ``{Most general AdS$_{3}$ boundary conditions},''
  {\em JHEP} {\bf 10} (2016) 023,
\href{http://www.arXiv.org/abs/1608.01308}{{\tt 1608.01308}}.

\bibitem{Grumiller:2017sjh}
D.~Grumiller, W.~Merbis, and M.~Riegler, ``{Most general flat space boundary
  conditions in three-dimensional Einstein gravity},'' {\em Class. Quant.
  Grav.} {\bf 34} (2017), no.~18, 184001,
  \href{http://www.arXiv.org/abs/1704.07419}{{\tt 1704.07419}}.

\bibitem{Grumiller:2022qhx}
D.~Grumiller and M.~M. Sheikh-Jabbari, {\em {Black Hole Physics: From Collapse
  to Evaporation}}.
\newblock Grad.Texts Math. Springer, 11, 2022.

\bibitem{Henneaux:2021yzg}
M.~Henneaux and P.~Salgado-Rebolledo, ``{Carroll contractions of
  Lorentz-invariant theories},'' {\em JHEP} {\bf 11} (2021) 180,
  \href{http://www.arXiv.org/abs/2109.06708}{{\tt 2109.06708}}.

\bibitem{Andringa:2010it}
R.~Andringa, E.~Bergshoeff, S.~Panda, and M.~de~Roo, ``{Newtonian Gravity and
  the Bargmann Algebra},'' {\em Class. Quant. Grav.} {\bf 28} (2011) 105011,
  \href{http://www.arXiv.org/abs/1011.1145}{{\tt 1011.1145}}.

\bibitem{Perez:2021abf}
A.~P\'erez, ``{Asymptotic symmetries in Carrollian theories of gravity},'' {\em
  JHEP} {\bf 12} (2021) 173, \href{http://www.arXiv.org/abs/2110.15834}{{\tt
  2110.15834}}.

\bibitem{Perez:2022jpr}
A.~P\'erez, ``{Asymptotic symmetries in Carrollian theories of gravity with a
  negative cosmological constant},'' {\em JHEP} {\bf 09} (2022) 044,
  \href{http://www.arXiv.org/abs/2202.08768}{{\tt 2202.08768}}.

\bibitem{Aviles:2025ygw}
L.~Avil\'es, O.~Fuentealba, D.~Hidalgo, and P.~Rodr\'\i{}guez, ``{AdS$_3$
  Carroll gravity: asymptotic symmetries and C-thermal configurations},''
  \href{http://www.arXiv.org/abs/2503.18818}{{\tt 2503.18818}}.

\bibitem{Ecker:2024czh}
F.~Ecker, A.~Fiorucci, and D.~Grumiller, ``{Tantum gravity},'' {\em Phys. Rev.
  D} {\bf 111} (2025), no.~2, L021901,
  \href{http://www.arXiv.org/abs/2501.00095}{{\tt 2501.00095}}.

\bibitem{Grumiller:2007ju}
D.~Grumiller and R.~McNees, ``Thermodynamics of black holes in two (and higher)
  dimensions,'' {\em JHEP} {\bf 04} (2007) 074,
\href{http://www.arXiv.org/abs/hep-th/0703230}{{\tt hep-th/0703230}}.

\end{thebibliography}
\end{document}